\documentclass[3p,times,12pt]{elsarticle}

\usepackage{amssymb}
\usepackage{amsmath}
\usepackage{hyperref}
\usepackage{array}
\usepackage[table]{xcolor}
\usepackage{xurl} 
\usepackage{siunitx}
\usepackage{subfigure}
\usepackage{multirow}
\usepackage{soul}
\usepackage{float}

\graphicspath{{figs/}}

\hypersetup{pdfauthor=KDSource} 

\journal{Annals of Nuclear Energy}

\begin{document}

\begin{frontmatter}

\title{KDSource, a tool for the generation of Monte Carlo particle sources using kernel density estimation}

\author[a]{N. S. Schmidt}
\author[a]{O. I. Abbate}
\author[a]{Z. M. Prieto\corref{cor1}}
\ead{zoe.prietom@gmail.com}
\cortext[cor1]{Corresponding author}
\author[a,b]{J. I. Robledo}
\author[c]{J. I. Márquez Damián}
\author[a]{A. A. Márquez}
\author[a,b]{J. Dawidowski}

\address[a]{Comisi\'on Nacional de Eneg\'ia
  At\'omica, Universidad Nacional de Cuyo, Argentina}

\address[b]{Consejo Nacional de Investigaciones Cient\'ificas y
  T\'ecnicas (CONICET), Argentina}

\address[c]{European Spallation Source ERIC, Sweden}

\begin{abstract}

  Monte Carlo radiation transport simulations have clearly contributed
  to improve the design of nuclear systems. When performing in-beam or
  shielding simulations a complexity arises due to the fact that
  particles must be tracked to regions far from the original source or
  behind the shielding, often lacking sufficient statistics.
  Different possibilities to overcome this problem such as using
  particle lists or generating synthetic sources have already been
  reported. In this work we present a new approach by using the
  adaptive multivariate kernel density estimator (KDE) method. This
  concept was implemented in \texttt{KDSource}, a general tool for
  modelling, optimizing and sampling KDE sources, which provides a
  convenient user interface. The basic properties of the method were
  studied in an analytical problem with a known density
  distribution. Furthermore, the tool was used in two Monte Carlo
  simulations that modelled neutron beams, which showed good agreement
  with experimental results.
\end{abstract}


\end{frontmatter}

\section{Introduction}
\label{sec:intro}

The calculation of radiation beams is a specific topic of the general
radiation transport problem. Normally it is decoupled from
the source since the nuclear reactions that govern the generation of
particles in the source are independent of the specific interactions
that take place in the beam path. In the design of experimental
nuclear facilities, radiation beams are usually transported far away
from the source both to reduce the background signal in the
measurements and the radiation dose of the personnel. When trying to
evaluate the beam under different operating conditions, it is useful to
have a source that can be re-sampled. A possible solution is to
capture the particles that are produced in the position of the source
by fission, decay or other events, and transport them into the beam
\cite{jazbec2021dose}. Although this is the most common approach, its
application to Monte Carlo simulation requires complex variance
reduction techniques (such as specifically generated weight windows
using CADIS and FW-CADIS \cite{mosher2013advantg}), and has a high
computational cost. Another approach is to use particle lists or track
files, in which the particles that cross a specific surface are
recorded and reused in a downstream simulation
\cite{kittelmann2017monte}. This method is computationally more
efficient but it may create artificial hot spots in the final solution
due to the discrete nature of the recorded data. To avoid this
problem, a third solution is to use a synthetic source, fitting an
analytical distribution for each of the source variables (energy,
position, direction and time) \cite{willendrup2020mcstas,
  ersez2018validation}.  Although this solution is simple, it strongly
depends on the expertise or knowledge of the user and usually fails to
capture implicit (or less evident) correlations between variables
\cite{fairhurst2017calculo, ayala2019implementacion}. With this in
mind we propose the use of the adaptive multivariate Kernel Density Estimation
(KDE) method to estimate the source distribution at a given point in
the beam trajectory, which seeks to overcome the discussed limitations of the
previous approaches. Similar ideas have been proposed by Tyagi et
al. in medical physics applications \cite{tyagi2006proposed},
generating phase-space samples without the need of direct particle
simulation, and by Banerejee and Burke \cite{banerjee2010kernel,
  burke2016kernel} for the improvement of tally computations. However,
its use in source generation has not been yet implemented in a
practical way in any of the main beam simulation codes.

The aim of this work is to present \texttt{KDSource}, a tool for
general modeling of sources by means of adaptive multivariate kernel
density estimation, especially suited for radiation beam and radiation
shielding simulations. The software consists of a module for KDE model
optimization, and another for sampling (i.e. generating new particles
using the previously optimized model).

The organization of this paper is the following. For those unfamiliar
with KDE, section \ref{theo} provides a brief introduction to the
theoretical concepts used. Section \ref{sec:implementation} presents
\texttt{KDSource}, the proposed tool for modeling and sampling source
distributions by means of KDE.  The results are summarized in three
parts. In section \ref{sec:Verification} \texttt{KDSource} is used to
estimate a known probability distribution of a sample, to verify the
goodness of its fit, and to study its convergence with the number of
particles in the sample. Section \ref{sec:validation} shows the comparison
between measurements performed on a pulsed neutron source based on a
linear accelerator and its simulation using \texttt{KDSource}.
Finally, in section \ref{sec:application}, an example of the
application of \texttt{KDSource} to a problem of neutron radiation
shielding in a neutrography facility in a research reactor is shown.
A general discussion about the advantages of using this tool,
comments on the results obtained, and the conclusions of this work
are presented in section \ref{sec:conclusions}.

\section{Theory}
\label{theo}

Kernel density estimation is a non-parametric method to estimate an
unknown probability density function (PDF) using a sample
set. Although many other non-parametric methods exist, such as
histograms or frequency polygons, KDE has the advantage that it is
independent of the bin size and the starting bin, and that it produces
a smooth estimate of the PDF, allowing for a better representation of
multimodality \cite{scott2015multivariate}. In KDE, the goodness of
the PDF estimation depends crucially on two concepts: the chosen
kernel function and the bandwidth size. Multivariate kernel density
estimation is the extension of KDE to higher dimensions, when the
joint probability distribution of a set of variables must be
estimated. On the other hand, adaptive KDE refers to the possibility
of using a different smoothing parameter in different regions of the
PDF domain.  The name KDE will refer to all variants of the kernel
density estimation method throughout this work.

To give a brief introduction to the method, let us consider a problem
characterized by a joint PDF of $D$ variables
$f({x}_{1},{x}_{2},...,{x}_{\rm D})$  of the $D$-dimensional phase
space. Consider $N$
observations $\textbf{p}_1,\textbf{p}_2,...,\textbf{p}_{\rm N}$,  where each $\textbf{p}_{\rm i}=((p_{i})_1,(p_{i})_2,...,(p_{i})_{D}$) is a $D$-dimensional vector of the phase
space. Each of the $D$ variables is characterized by their
standard deviations $\sigma_1,\sigma_2,...,\sigma_{\rm D}$.  When
dealing with any physical problem the components of the vector
$\textbf{p}$ can have different units (energy, length, angle, etc.)
and can vary by different orders of magnitude, so a good
practice that will be used in this work consists in rescaling them,
dividing them by their corresponding standard deviation. The result is
that the new variables will be dimensionless. Let the rescaled variables be
$\tilde{\textbf{p}}_1,\tilde{\textbf{p}}_2,...,\tilde{\textbf{p}}_{\rm
  D}$, the KDE method estimates the PDF by setting a specified kernel
function at the exact position of each observation, and adding these
functions over all data points (with a corresponding normalization
factor).  The multivariate KDE method estimates the unknown joint PDF
$f(x_{1},x_{2},...,x_D)$ from
the $N$ observations using the estimator $\hat{f}(x_{1},x_{2},...,x_D)$ given by
\begin{equation}
\hat{f}(\textbf{x})=    \hat{f}(x_{1},x_{2},...,x_{D})=\sum_{\rm i=1}^{N} w_{\rm i} \left
      \{\prod_{\rm j=1}^{D}\frac{1}{ h}
      K\left(\frac{x_{\rm j}-(\tilde{p}_{\rm i})_{\rm j}}{h}\right) \right \} 
    \label{eq:multiD}
\end{equation}
where
$\frac{1}{h}K(\frac{x_{\rm j}-(\tilde{p}_{\rm i})_{\rm j}}{h}) \equiv
K_{h}(x_{\rm j},(\tilde{p}_{\rm i})_{\rm j})$ is a user-specified
kernel function, $w_{\rm 1},w_{\rm 2},...,w_{\rm N}$ is a set of
normalized weights giving the observations different importance in the
estimation if needed (if not, $w_i=1/n$ for all $i$), and $h$ is a
hyper-parameter of KDE called the bandwidth or smoothing parameter
\cite{Silverman86}. This hyper-parameter may be interpreted as the
standard deviation of the selected kernel. The definition in
Eq. (\ref{eq:multiD}) has been simplified with respect to other
definitions given in the literature by taking the multivariate kernel
as the product of $D$ one-dimensional kernels, and replacing the
bandwidth matrix $H$ by $h^2 I$ (where $I$ is the identity matrix).

In KDE, the kernel can be any normalized and positive definite
function. Even though there are no constraints on symmetry properties,
symmetric kernels are usually preferred. Some of the most commonly
used are Epanechnikov, Gaussian, triangular, Tophat and exponential
kernels, each of which is better suited in different scenarios,
and some authors go further by carefully choosing the
kernel in order to reduce the contribution of the bias to the Mean
Integrated Square Error (MISE) \cite{Bartlett1963}. The bandwidth $h$ is the
most important parameter to define. If $h$ is very small, KDE will
reproduce the statistical noise, due to the finite size of the
sample. On the contrary, if $h$ is too wide, it will generate an
over-smoothing representation of the PDF, and it can produce a loss of
information regarding the underlying physics. Therefore, there is a
compromise between overfitting and over-smoothing when defining the
optimal bandwidth. There are many methods to select an optimal
bandwidth by minimizing an estimator of the total error. A common rule
of thumb for bandwith selection is Silverman's rule \cite{silverman},
which assumes that the data are distributed normally so it only works
well in unimodal distributions.

As an alternative to conventional KDE where a single value of $h$ is
chosen, a generalization of the method, called adaptive KDE, allows a
different bandwidth value for each data point. In such cases,
Eq.(\ref{eq:multiD}) is not strictly valid. To avoid confusions, the
reader interested in observing how this equation is modified can refer
to Eq. (6.84) in Ref. \cite{scott2015multivariate}. In adaptive KDE the
value of the bandwidth depends on each data point by a defined
criterion \cite{AKDE}. The one of interest to this work is to use the
k-Nearest Neighbor (kNN) algorithm to select the bandwidth $h_i$ for
each data point $\mathbf{x}_i$, taking $h_i$ as the distance to the
$k$-th nearest neighbor of the data point.

Another concept that we will use in this work is the Kulback-Leibler
divergence \cite{KLdivergence}

\begin{equation}
    D_{KL}(P||Q) = \int_{-\infty}^{\infty} P(x) \log\left(\frac{P(x)}{Q(x)}\right) dx
\end{equation}
used to compare the estimated distribution $Q(x)$ to the known
analytical distribution $P(x)$. This divergence is the expectation
value of the logarithmic difference between the probability densities
$P(x)$ and $Q(x)$, where the expectation value is taken with respect to the
probability $P(x)$. This measure is 0 if both distributions are equal,
and results in a positive value if the distribution $Q(x)$ differs
from $P(x)$. In information theory it is called the relative entropy
and is interpreted as the amount of information lost when $Q$ is used
to approximate $P$ (or as the statistical distance between $P(x)$ and $Q(x)$).

\section{KDSource: a tool for modeling and sampling sources}
\label{sec:implementation}

\texttt{KDSource}, the tool presented in this work, uses the adaptive
multivariate kernel density estimator to give an estimate of the
underlying joint probability distribution of the phase space variables
by means of a particle list recorded in an intermediate position in a
Monte Carlo simulation, and allows sampling from the estimated
distribution.  The tool consists of a Python application programming
interface (API) for the modeling, analysis, and plotting of the
source, and a C API for sampling the estimated distribution that
provides compatibility with Monte Carlo codes for sampling sources
on-the-fly. It also includes a command line tool with a high level
interface for particle sampling and access to several helpful
utilities. The source code can be found in the \texttt{KDSource}
GitHub repository \cite{KDSource2021}.  \texttt{KDSource} uses the
\texttt{MCPL} \citep{MCPL} format for particle lists, which is a
general binary format that can easily be used as input or output for
codes such as \texttt{MCNP} \cite{MCNP5, MCNPX, MCNP6.2},
\texttt{PHITS} \cite{PHITS}, \texttt{GEANT4} \cite{Geant4},
\texttt{McStas} \cite{McStas1, McStas2, McStas3}, \texttt{McXtrace}
\cite{McXtrace} and, in the modified version included in the
\texttt{KDSource} package, \texttt{TRIPOLI-4} \citep{TRIPOLI} and
\texttt{OpenMC} \citep{romano2013openmc}. The following subsections
describe different parts of \texttt{KDSource} and its implementation.

\subsection{Data preprocessing}
\label{sec:preproc}
This part consists of two steps: specific transformations of variables
and a subsequent normalization. The first step starts from the phase
vector $\textbf{p}$, which together with the weight $w$ comprises the
list of parameters defining the state of a particle
\begin{equation}
  \textbf{p} = (E, x, y, z, d_x, d_y, d_z),
  \label{pmcpl}
\end{equation}
where $E$ is the particle energy, $(x,y,z)$ are its position Cartesian
coordinates, and $(d_x,d_y,d_z)$ its flight direction
unit-vector. Variables time and polarization, which also can be taken
into account, are ignored in this work.  The phase vectors, recorded
as a list in MCPL format, make up the track lists that record the
trajectory of the particles. They must contain only one type of
particles (neutrons, photons, etc.). Depending on the source geometry
and other specific characteristics of the problem, the phase vector as
presented in (\ref{pmcpl}) may not be appropriate for the problem.
\texttt{KDSource} makes the necessary variable changes, and selection
of variables relevant to the problem. After this process, the
\texttt{MCPL} format (\ref{pmcpl}) can be altered, and hence the
vector $\textbf{p}$ in the KDE model shown in Eq. (\ref{eq:multiD})
will correspond to the transformed variables.  Those transformations
can be chosen from the set of implemented functions.  Thus, the energy
can be transformed into the variable lethargy $u=\log(E_0/E)$ ($E_0$
being a reference energy), suitable for cases in which neutron
thermalization processes are relevant.  Regarding the position, flat
sources can be modeled using only the $x,y$ components of the
position. Neutron guide sources, which are tube-shaped surface sources
with a rectangular cross-section, are also implemented. Finally,
either polar coordinates (variables $\theta$ and $\varphi$) or
Cartesian coordinates are used for direction variables.  It is also
possible to leave any of the variables untransformed (i.e., in the
starting \texttt{MCPL} format).

Since, as already mentioned, the variables that intervene in the
problem to be treated have different units and orders of magnitude, a
previous step to the application of the method consists of
renormalizing them by their respective standard deviations. In such a
way we work with normal dimensionless variables.  Furthermore, by
using a scaling factor different from $\sigma_j$ for variable $j$, the
quality of the estimation along that dimension can be controlled. In
particular, an importance factor (or degree of priority) $\alpha_j$
for each variable $j$ can be fixed by setting its scaling factor as
$\sigma_j/\alpha_j$. An $\alpha_j$ higher than 1 leads to a better
estimation of the distribution along the variable $j$, at the expense
of a decrease in the quality on the other variables.

\subsection{Modeling}
\label{sec:modeling}

At the moment, only Gaussian kernels are available in
\texttt{KDSource}, but a future release will also allow the
possibility to select different kernel functions. The bandwidth
parameter can be selected by three different means, namely using
Silverman's Rule, minimizing a total error estimator, or using
adaptive KDE, which are described below.
\begin{enumerate}
\item Silverman's rule \cite{silverman} is used to determine the value
  of $h$ in the kernel function assigned to each data point.

\item The Maximum-Likelihood Cross-Validation (MLCV) method
  \cite{MLCV} with $k$-fold cross validation \cite{Refaeilzadeh2016}
  ($k=10$) is used to select the optimal bandwidth $h$, which is
  assigned to every data point.  The method consists first in dividing
  the data into equally sized k segments or folds. Then, one of the
  folds is left for testing (with $N_{test}=N_{fold}$ particles) and
  the remaining $k-1$ folds are used for training (with
  $N_{train}=(k-1) N_{fold}$). The training split is used to estimate
  the density with KDE, and the test split is used to evaluate the
  estimated density in a quantity called Figure of Merit (FoM). This
  procedure is repeated $k$ times, each time selecting a different
  fold as test, and the average FoM over all folds is finally
  calculated. This process is done for different bandwidths and the
  one that maximizes the averaged FoM is selected as the optimal
  one. There are many different FoMs possible to choose. The MLCV
  method is based on maximizing the following FoM:

\begin{equation}
    \mathrm{FoM}_{\mathrm{MLCV}} = \sum_{i=1}^{N_{\mathrm{test}}} w_i \log(\hat{f}_{\mathrm{train}}(\textbf{x}_i))
    \label{eq:MLCV}
\end{equation}
where the probability density $\hat{f}_{\mathrm{train}}$ is estimated
with KDE using the training split of the data and afterwards is
evaluated in the test split. Each data point of the test split is
weighed with the weight $w_i$ already mentioned in Sect
\ref{sec:preproc}.

\item Adaptive KDE employing the kNN algorithm as described in the
  previous section is used to select a seed bandwidth $h_i$ for the
  kernel function assigned to each data point
  $\mathbf{x_i}$. Subsequently, the vector $\vec{h}$ is multiplied by
  a scale factor that is optimized using the MLCV method. This case is
  useful when the distribution to be estimated varies by several
  orders of magnitude between two regions of interest in phase space,
  and is generally the recommended bandwidth selection method.
\end{enumerate}

\subsection{Sampling}

The sampling stage is performed once the estimation of the
distribution described above is completed. \texttt{KDSource} samples
the estimated distribution by applying multidimensional perturbations
$\delta$ on the transformed sample data. These perturbations are
distributed according to the selected kernel function and have a
dispersion given by the chosen optimal bandwidth $h$ ($h_i$ in the
case of adaptive KDE where the sample data $\mathbf{x_i}$ are
perturbed), i.e.,

\begin{equation}
\widetilde{\textbf{x}}=\textbf{x}_i+ \delta ,\: \: \: \: \: \: \: \: \: \:   \delta \sim K_{h}= N(0,h),
\label{eq:sample}
\end{equation}
where $\textbf{x}_i$ is a data point from the complete list of
simulated data, which is sampled following the list order, and
$\delta$ has a distribution according to $K_h$, which is normal with
mean value 0 and variance $h$. The sample generated by this method
follows the estimated distribution given in Eq. (\ref{eq:multiD})
\cite{scott2015multivariate}. Finally, the generated phase vectors
$\widetilde{\textbf{x}}$ are reconverted to the \texttt{MCPL} format,
to be used as input of a new simulation.

\subsection{Workflow and implementation}

\begin{figure}[htbp!]
  \centering \includegraphics[width=.7\textwidth]{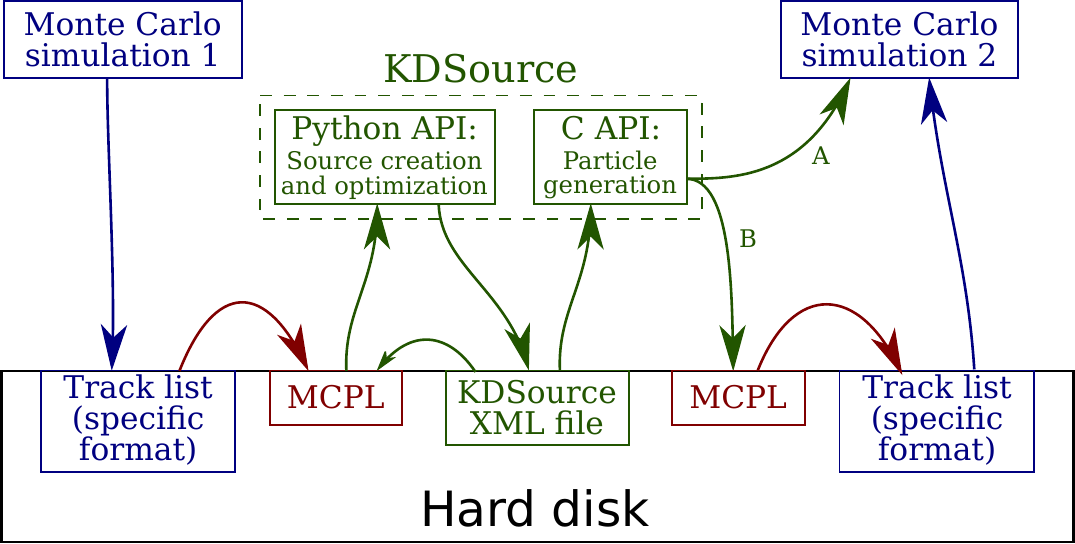}
  \caption{Workflow for the \texttt{KDSource} algorithm. Blue
    indicates the files and actions related to the Monte Carlo
    simulation codes, red indicates the \texttt{MCPL} files and
    related conversions, and green indicates the files and actions
    related to the \texttt{KDSource} tool.}
    \label{fig:workflow}
\end{figure}

The \texttt{KDSource} workflow is depicted in
Fig. \ref{fig:workflow}. It begins with an initial Monte Carlo
simulation, in which a track list is recorded at an intermediate
position between the original source (e.g., the reactor core) and the
point of interest (e.g., a beam end). Since the track list has the
specific format of the Monte Carlo code, the first step is to convert
it to \texttt{MCPL} format. The particle list is then loaded from
Python, and used to fit a KDE model with the \texttt{KDSource} Python
API (as explained in section \ref{sec:modeling}), which is then saved
as an XML file containing the optimized bandwidth, the path to the
\texttt{MCPL} file, and other source parameters. The C API then loads
this file to rebuild the KDE model and use it to generate new
particles. These particles can be either directly inserted into the
second simulation (path A in Fig. \ref{fig:workflow}) as an
``on-the-fly source'', or saved in a new \texttt{MCPL} file, usually
larger than the first one (path B in Fig. \ref{fig:workflow}), which
can then be transformed to the required specific format and used as
input for the second simulation. It should be noted that the Monte
Carlo codes used in this workflow must be \texttt{MCPL}
compliant. These codes were listed at the beginning of this
section. So far, on-the-fly sources were only implemented for
\texttt{McStas} and \texttt{TRIPOLI-4}. Although each KDE source can
produce only one type of particle (neutron, photon, etc.), two or more
sources that produce different types of particles can be overlapped in
the second simulation, allowing coupled neutron-photon simulations
with the \texttt{KDSource} methodology.

In the Python API, KDE is implemented using the \verb|TreeKDE| class
of \verb|KDEpy| library \citep{KDEpy}. The main advantage of this
library is that it implements adaptive KDE, and that the class
\verb|TreeKDE| includes acceleration via the KDTree structure
\citep{KDTree}, which reduces the computational cost of evaluating the
estimated density and optimizing the KDE source.
The proposed tool also uses some utilities from \verb|Scikit-Learn|
\citep{sklearn} (KNN search, K-folding), and CPU parallelization via the 
\verb|joblib| library \citep{joblib}.

\section{Results}

This section is divided into three parts. In the first one,
\texttt{KDSource} is tested with a known \textit{a priori} analytical
distribution of correlated variables. To do so, first an analytical
multivariate joint distribution is constructed and sampled, after
which \texttt{KDSource} is applied to obtain the estimated
distribution. Then, both distributions are compared using the
Kulback-Leibler divergence. In the subsequent parts, examples of
application to the analysis of simulations of real cases are
presented. The second part, shows an application of \texttt{KDSource}
in a Monte Carlo neutron transport problem to reproduce the spectrum
obtained at the LINAC neutron source at CAB (Bariloche, Argentina). In
this part, \texttt{KDSource} is used to estimate the joint
distribution of the phase-space vector at a certain distance from the
source, and then it is sampled to continue the simulation at a further
distance. Here it is clear how \texttt{KDSource} is an advantageous
tool to accelerate calculations at a great distance between the
neutron source and the detector, with accurate results.  Finally, in
part three, \texttt{KDSource} is used to enhance the convergence of
the dose rate map in a neutron tomography facility in the RA-6
research reactor at CAB.

\subsection{Verification}
\label{sec:Verification}

In order to evaluate the performance of \texttt{KDSource}, a benchmark
test was performed. For this purpose, we generated a multivariate
analytical distribution representing a neutron source, with which the
variables were sampled, and the KDE method was applied to generate an
estimated PDF.  The calculation details of this process are described
in a Google Colab notebook in the GitHub repository
\cite{KDSource2021}.

We chose a 2D-flat source, whose variables (see section
\ref{sec:implementation}) are the lethargy $u$, the positions $x,y$,
and the direction vector
$(d_x,d_y,d_z)= (\sin\theta\cos\varphi, \sin\theta\sin\varphi,
\cos\theta)$.  The distribution function was constructed so that
there were correlated variables, in order to evaluate the ability of
the tool to correctly describe it.

The joint distribution function was defined as
\begin{equation}
  \label{eqf}
  f(u,x,y,\mu,\varphi) = \left[ f_{u,1}(u)f_{x,1}(x) + f_{u,2}(u)f_{x,2}(x) \right] f_y(y) f_{\mu}(\mu) f_{\varphi}(\varphi),
\end{equation}
where $\mu=\cos\theta$, $f_{u_1}$, $f_{u_2}$, $f_{x_1}$, $f_{x_2}$, and $f_y$
are normal distributions, $f_{\mu}$  is linear, and $f_{\varphi}$ a
constant distribution, as defined below
\begin{gather}
  \label{deff}
f_{u,i}(u)= \frac{1}{\sqrt{2\pi}\sigma_{\rm u}}
  \exp{\left[-\frac{(u-\mu_{u,i})^2}{2\sigma_{\rm u}^2}\right]}\\
f_{x,i}(x)= \frac{1}{\sqrt{2\pi}\sigma_{\rm x}}
  \exp{\left[-\frac{(x-\mu_{x,i})^2}{2\sigma_{\rm x}^2}\right]}\\
f_{y}(y)= \frac{1}{\sqrt{2\pi}\sigma_{\rm y}}
  \exp{\left[-\frac{y^2}{2\sigma_{\rm y}^2}\right]}\\
f_{\mu}(\mu)= 2\mu \\
f_{\varphi}(\varphi)= \frac{1}{2\pi},
\end{gather}
where $u$, $x$ and $y$ $\in [-\infty,\infty]$, $\mu>0$ and $\phi \in
[0, 2\pi]$. The parameters employed in the Gaussian functions are summarized in
Table \ref{tab1}.
\begin{table}[h]
\centering
\begin{tabular}{|c|c|c|c|c|c|}
\hline
Parameter & $f_{u,1}$ & $f_{u,2}$ &  $f_{x,1}$  &  $f_{x,2}$ &  $f_{y}$  \\
\hline
 $\mu$  &  5  & 9  & 10  & -10 & 0 \\
 $\sigma$   &  1  & 1  & 10  &  10 & 10 \\
\hline
\end{tabular}

\caption{Parameters employed for the definition of the Gaussian
  functions in Eq.(\ref{eqf}).}
\label{tab1}
\end{table}
From the definition of the distribution function, we observe that the
lethargy variable $u$ is correlated with the position variable
$x$ (the distribution presents two non-aligned Gaussian peaks in the
$u-x$ plane). The remaining variables are independent.

\begin{figure}[htbp!]
    \centering
    \includegraphics[width=.5\textwidth]{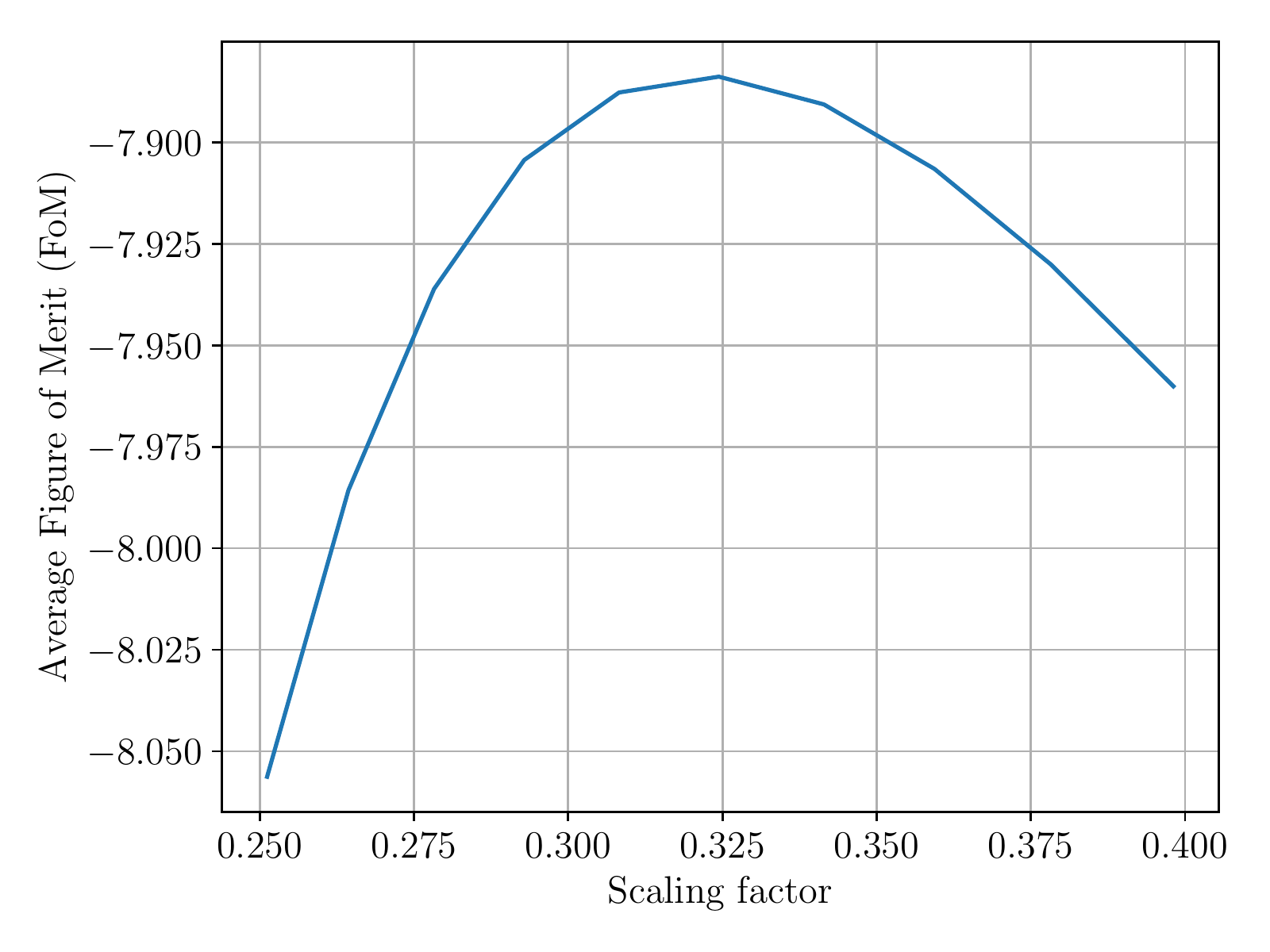}
    \caption{Average Figure of merit of the Maximum Likelihood Cross
      Validation method, as a function of the scaling factor of the
      seed bandwidth. The selected normalized bandwidth is the one
      with the highest average FoM value.}
    \label{fig:mlcv-h}
\end{figure}

\begin{figure}[H]
  \centering \subfigure{%
    \label{fig:KLD-conv}%
    \includegraphics[width=0.4\textwidth]{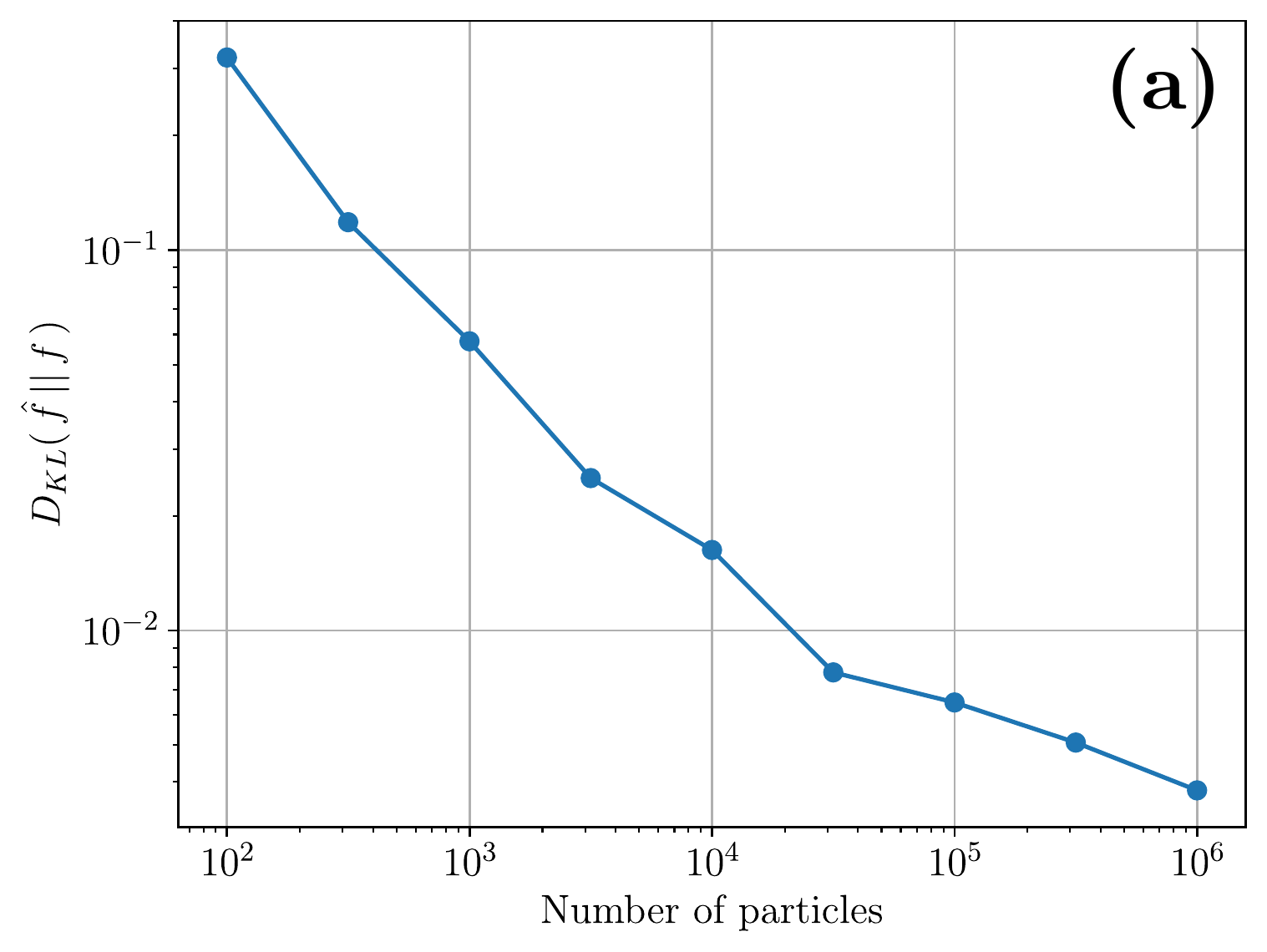}%
  } \hspace{.02\textwidth} \subfigure{%
        \label{fig:E-comp}%
        \includegraphics[width=0.4\textwidth]{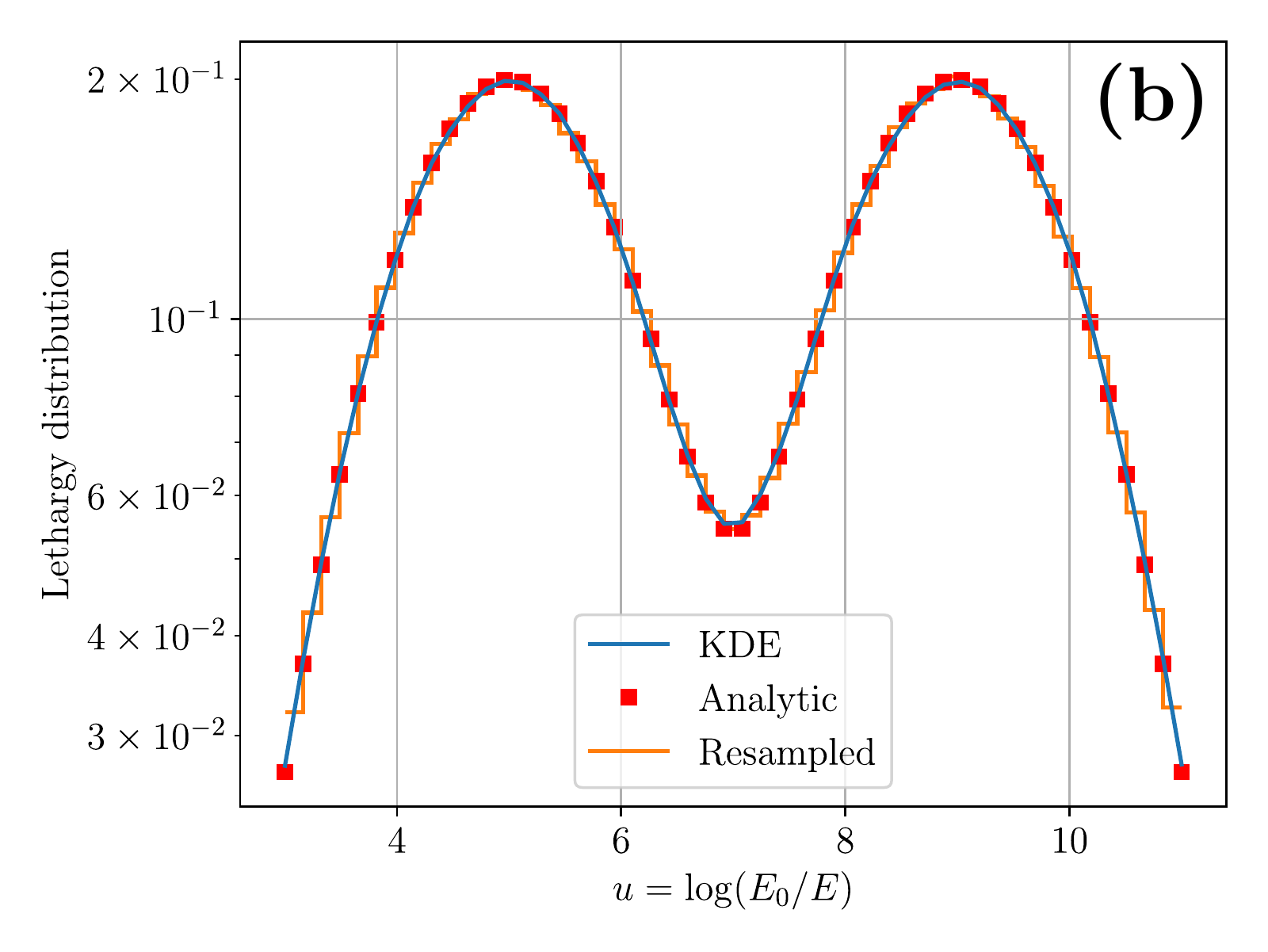}%
    }
    \caption{(a) Comparison between analytical and estimated energy
      spectra. Kulback-Leibler divergence between the analytical and
      estimated lethargy distribution, as a function of the number of
      training particles. (b) Comparison between analytical
      lethargy distribution, estimated distribution (with $N=10^6$)
      and histogram of the resampled particles.}
    \label{fig:E-conv}
\end{figure}

A random particle list with the distribution given by Eq. (\ref{eqf})
was generated using the \texttt{numpy} library within the Python
environment. From this list, the KDE method was applied using
\texttt{KDSource} with an optimized bandwidth through the kNN and MLCV
methods described in Sect. \ref{sec:modeling}.  The process was
repeated varying the number of particles in the list $N$. In
Fig. \ref{fig:mlcv-h} the MLCV figure of merit (Eq. \ref{eq:MLCV}) is
shown as a function of the scaling factor of the $h$ vector obtained
by kNN, for $N=10^6$ training particles, where a clear maximum that
defines the optimal bandwidth can be observed. Analytical and
estimated distributions were compared using the Kulback-Leibler
divergence ($D_{KL}$, see section \ref{theo}) for different number of
sampled particles, and the results are shown in
Fig. \ref{fig:KLD-conv}.  It can be seen that the $D_{KL}$ is a
decreasing function of $N$, meaning that the larger the sample set,
the closer is the estimated density to the true PDF. Nevertheless, the
information gain decreases with $N$. In fact, the $D_{KL}$ decrease
from $10^5$ to $10^6$ particles is around the 27\% of the decrease
from $10^4$ to $10^5$, and the 1\% of the same amount from $10^2$ to
$10^3$.

\begin{figure}[htbp!]
    \centering
    \includegraphics[width=.7\textwidth]{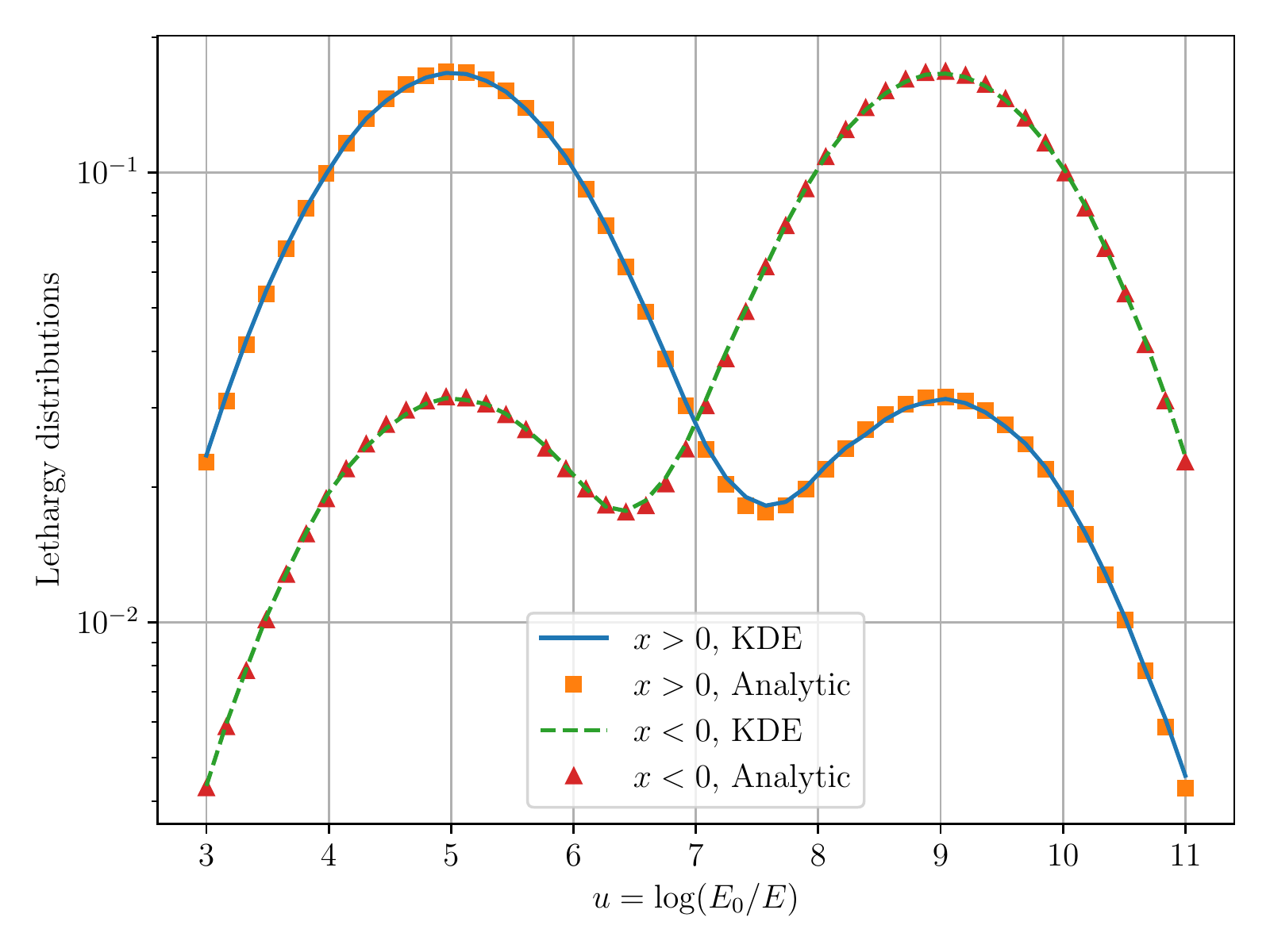}
    \caption{Comparison of the estimated and analytic energy
      distributions for $x<0$ (red triangles) and $x>0$ (orange
      squares).}
    \label{fig:E-x-correl}
\end{figure}

Fig. \ref{fig:E-comp} shows a comparison between the analytical lethargy
distribution (red squares), the \texttt{KDSource} estimated
distribution with $N=10^6$ (blue connected lines), and the histogram
obtained by sampling $10^6$ new particles from the estimated
distribution (orange steps). This figure manifests that the
distribution of the sampled points follows accurately the original
analytical distribution in the entire energy range. Moreover, we
checked the adequate modeling of the distribution correlation between
energy and $x$ by plotting the energy spectrum for different ranges of
$x$ to see its variation. In Fig. \ref{fig:E-x-correl} the estimated
and analytical distributions are compared for $x<0$ and $x>0$, showing
good agreement.

The results presented in this section show that the \texttt{KDSource}
tool successfully estimates an analytical correlated joint
distribution of the source's phase space vector, and that it allows to
generate new samples from it maintaining the correlations between
variables. In the following section, \texttt{KDSource} will be used to
produce particles in a Monte Carlo simulation.

\newpage
\subsection{Validation}
\label{sec:validation}

\begin{figure}[H]
  \includegraphics[width=\textwidth]{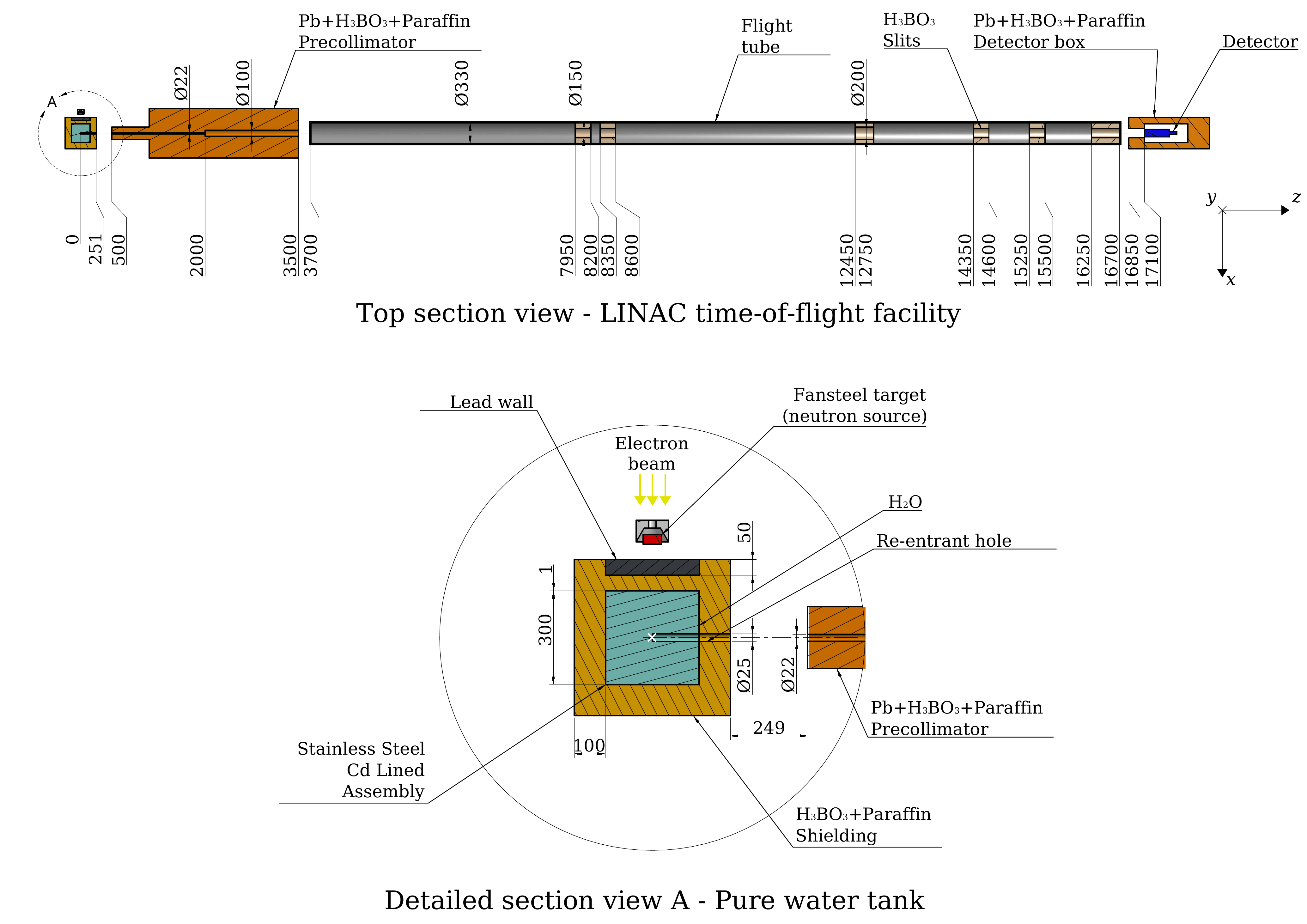}
    \caption{Experimental scheme for the measurement of neutron
      spectra in pure H$_2$O. In the upper part a general sketch of
      the experimental facility is observed. Electrons from the LINAC (not
      shown in the figure) hit a Fansteel target where fast neutrons
      are generated. On the left is the moderator consisting of a
      water cubic container shielded with paraffin and boric acid.
      The neutrons are extracted through a re-entrant hole facing the
      flight tube and are detected at a distance of 17 m. On the right
      the detector placed into a shielding made of lead, boric acid
      and paraffin is shown. The lower graph shows a detail of the
      moderator system. The white cross in the lower graph indicates
      the common coordinate origin for the \texttt{OpenMC} and
      \texttt{McStas} simulations. All the dimensions shown in the
      graph are in mm.}
    \label{fig:LINAC-Geometry}
\end{figure}

Next, we will show the use of \texttt{KDSource} in computational
simulations applied to neutron spectra measurement experiments made
in a pulsed neutron source based on a linear electron accelerator
(LINAC). The experiments, described in detail in
Ref. \citep{abbate1974linac}, correspond to those carried out at the
LINAC facility of the Bariloche Atomic Center (CAB) during the
1970s.  They consist in the measurement of neutron spectra of pure
water at \SI{23}{\celsius} (\SI{296.15}{\kelvin}). A simplified sketch
of the geometry with its dimensions can be observed in
Fig. \ref{fig:LINAC-Geometry}.  Fast neutrons are generated by a
photonuclear reaction from the bremsstrahlung radiation of the
electrons, which, accelerated by the LINAC to an energy of 25 MeV, are
stopped by a Fansteel target. The neutrons thus produced are
moderated in a \SI{25}{mm} side H$_2$O cubic tank, with a
re-entrant hole facing a 17 m long flight tube, at the end of
which the neutrons are detected.

The H$_2$O cubic tank and its shieldings were modeled with the
\texttt{OpenMC} Monte Carlo code \citep{romano2013openmc} using the
ENDF/B-VII.0 nuclear data library \citep{chadwick2006endf}, and the
flight tube and its slits were modeled with the \texttt{McStas} Monte
Carlo code. The calculations were performed starting from a
neutron source, modeled as a \SI{3}{\centi\meter} radius
monodirectional flat disk with a Watt energy distribution
\begin{equation}
  \label{watt}
S(E)\:dE = c \, e^{-E/a} \sinh \sqrt{b\, E}\:dE,
\end{equation}
where $E$ is given in MeV, $a=$\SI{0.38}{\mega\eV},
$b=$\SI{7}{\mega\eV^{-1}}, and $c$ is a source particle normalization
constant set as \SI{1}{n/s}. All the particles that cross the surface
determined by the start of the re-entrant hole in the \texttt{OpenMC}
simulation were saved into an \texttt{HDF5} file named surface source
file (SSF), and only the ones that were directed to the flight tube
were filtered and saved into the \texttt{MCPL} format. The coordinate
origin common to the \texttt{OpenMC} and \texttt{McStas} simulations,
placed at the extreme of the re-entrant hole (center of the surface
source), is indicated with a white cross in the lower graph of
Fig. \ref{fig:LINAC-Geometry}. The $z$-axis is parallel to the
direction of the flight path.

In the experiments, the measured spectra were detected at
\SI{17}{\meter} from the source. This poses a problem for a
conventional Monte Carlo simulation since very few of the emitted
particles will reach a detector at such a great distance. A
conventional \texttt{OpenMC} simulation shows that out of
$1\times 10^{10}$ source particles, about $7\times 10^{6}$ cross the
reference surface with direction to the flight tube and, only $83$
reach the detector. To overcome this problem and increase the
statistics at the detector, \texttt{KDSource} was used to estimate the
SSF distribution generated by \texttt{OpenMC}.

Two sources with different lethargy importance factors $\alpha_{u}$
(see Section \ref{sec:preproc}) were estimated, resampled, and saved
into \texttt{MCPL} format to evaluate their impact in the energy
distribution at the end of the flight tube. For the first source (case
1) $\alpha_u$ was set to 1 (default value in \texttt{KDSource}) and
for the second one (case 2), $\alpha_u=10$.

The number of resampled particles was the same as in the original
\texttt{OpenMC} SSF, and the resampled particle weights were
normalized to preserve the \texttt{OpenMC} integral partial current
which is defined as
\begin{equation}
  \label{jn+}
    J_{\hat{\mathbf{n}}}^{+} = \dfrac{S_0}{N_\text{sim}} \sum_{i=1}^{N_\text{surf}} w_i,
\end{equation}
where $S_0$ is the simulation source factor, $N_\text{sim}$ is the
total number of source simulated particles, $N_\text{surf}$ is the
total number of particles that cross out the reference surface with
normal direction $\hat{\mathbf{n}}$, and $w_i$ are the particle
statistical weights. Also, following the notation used in
\cite{duderstadt1979transport}, the outgoing normal angular current
$j_{\hat{\mathbf{n}}}^{+}(\mathbf{r},E,\hat{\mathbf{\Omega}})$ can be
rewritten as
\begin{equation}
    j_{\hat{\mathbf{n}}}^{+}(\textbf{r},E,\hat{\mathbf{\Omega}})= J_{\hat{\mathbf{n}}}^{+}\:f(\textbf{r},E,\hat{\mathbf{\Omega}}),
\end{equation}
where $f(\textbf{r},E,\hat{\mathbf{\Omega}})$ is a joint PDF obtained
as a multivariable histogram weighted with the particle weights.

\begin{figure}[H]
    \centering
    \subfigure{
        \includegraphics[width=0.47\textwidth]{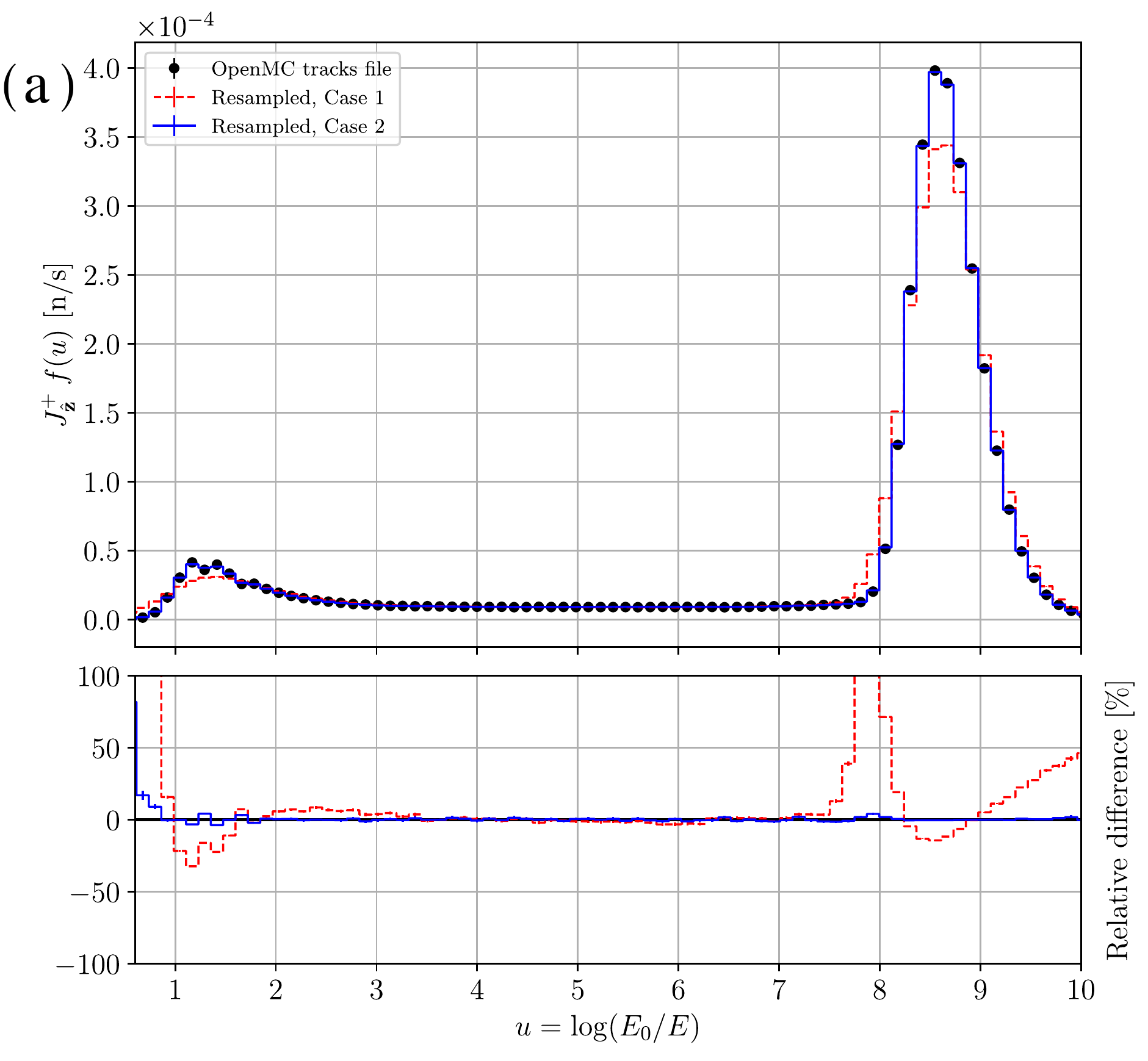}
        \label{subfig:J1-lethargy}
    }
    \subfigure{
        \includegraphics[width=0.48\textwidth]{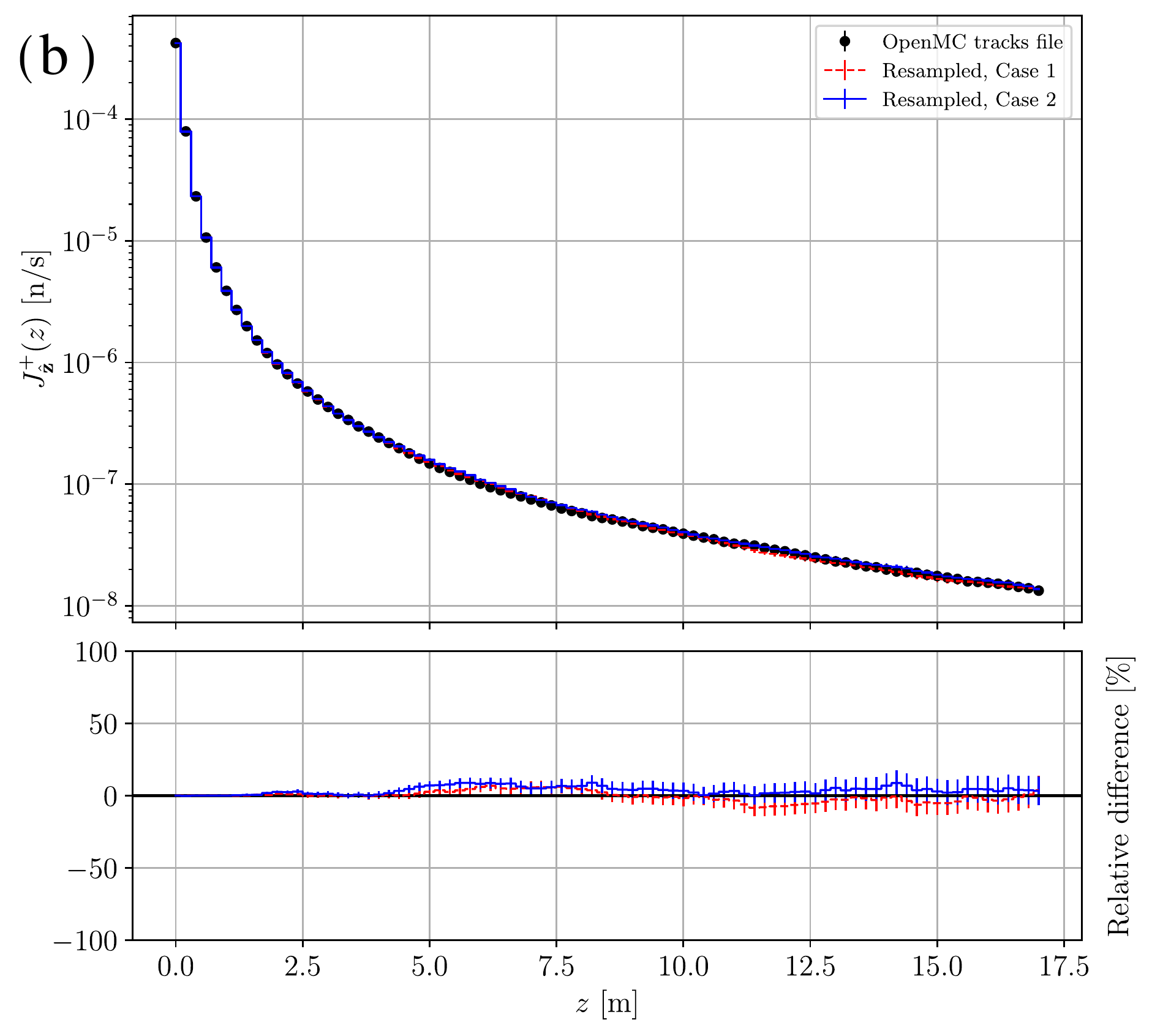}
        \label{subfig:J1-z}
    }
    \caption{Comparison between the partial current distributions and
      integral values obtained from the \texttt{OpenMC} SSF and the
      resampled ones with \texttt{KDSource} code for different
      lethargy scaling factors ($\alpha_{u}=1.0$ for case 1, and
      $\alpha_{u}=10.0$ for case 2). (a) Lethargy distribution at the
      origin of the flight path ($z=0$ m). (b) Integral partial
      current as a function of the flight length. The lower frames
      show the relative differences between cases 1 and 2.}
    \label{fig:J1-testing}
\end{figure}

The lethargy distributions $J_{\hat{\mathbf{z}}}^{+}f(u)$ at the
origin of the flight path ($z=\SI{0}{\meter}$) for both cases are
shown in Fig. \ref{subfig:J1-lethargy}, and the values of the integral
partial current as a function of the flight length $z$ are shown in
Fig. \ref{subfig:J1-z}. Both figures also show the relative difference
between the resampled sources and the \texttt{OpenMC} SSF. Also, a
summary with the integral values for different energy and angular
ranges is reported in Table \ref{table:J1-integral}. The large
relative differences shown for the first case can be attributed to the
high value of the second derivative in those lethargy regions, as
discussed in \citep{scott2015multivariate, stoker1993smoothing}.  The
importance factor used for the lethargy variable in the second case
reduces the broadening of the lethargy distribution, showing a good
agreement with the original SSF. This can also be observed from the
integral values shown in Table \ref{table:J1-integral}, since the
relative difference is larger in the first case than in the second one
for all the chosen ranges.

\begin{table}[H]
\centering
\resizebox{\textwidth}{!}{
\begin{tabular}{|c|c|ccccc|}
\hline
\multicolumn{2}{|c|}{\multirow{2}{*}{Integration range}}       & \multirow{2}{*}{OpenMC track file}  & \multicolumn{4}{c|}{Resampled}                                                                                          \\
\multicolumn{2}{|c|}{}                             &
                                                                                        &
                                                                                          \multicolumn{2}{c}{Case 1 ($\alpha_{u}=1.0$)}                                 & \multicolumn{2}{c|}{Case 2 ($\alpha_{u}=10.0$)}                                \\
  \hline
$\Delta E$ {[}eV{]}              & $\Delta \mu$   & $J^{+}_{\hat{\mathbf{z}}}$ {[}n/s{]} & $J^{+}_{\hat{\mathbf{z}}}$ {[}n/s{]} & Difference {[}\%{]} & $J^{+}_{\hat{\mathbf{z}}}$ {[}n/s{]} & Difference {[}\%{]} \\ \hline
Total                                & Total              & $4.226(2)\times 10^{-4}$             &                                      &                     &                                      &                     \\ \hline
\multirow{3}{*}{{[}1$\times 10^{-5}$, 0.3{]}} & Total              & $3.281(1)\times 10^{-4}$             & $3.265(1)\times 10^{-4}$             & $-0.47(6)$          & $3.280(1)\times 10^{-4}$             & $-0.01(6)$          \\
                                 & {[}0.0, 0.5{]} & $8.316(7)\times 10^{-5}$             & $8.268(7)\times 10^{-5}$             & $-0.6(1)$           & $8.369(7)\times 10^{-5}$             & $0.6(1)$            \\
                                 & {[}0.5, 1.0{]} & $2.449(1)\times 10^{-4}$             & $2.439(1)\times 10^{-4}$             & $-0.43(7)$          & $2.445(1)\times 10^{-4}$             & $-0.23(7)$          \\ \hline
\multirow{3}{*}{{[}0.3, 2$\times 10^7${]}}  & Total              & $9.45(1)\times 10^{-5}$              & $9.60(1)\times 10^{-5}$              & $1.6(1)$            & $9.46(1)\times 10^{-5}$              & $0.0(1)$            \\
                                 & {[}0.0, 0.5{]} & $3.134(6)\times 10^{-5}$             & $3.213(4)\times 10^{-5}$             & $2.5(2)$            & $3.198(4)\times 10^{-5}$             & $2.0(2)$            \\
                                 & {[}0.5, 1.0{]} & $6.319(8)\times 10^{-5}$             & $6.389(6)\times 10^{-5}$             & $1.1(2)$            & $6.258(6)\times 10^{-5}$             & $-1.0(2)$           \\ \hline
\end{tabular}
}
\caption{Comparison between integral $J^{+}_{\hat{\mathbf{z}}}$ values
  obtained from the \texttt{OpenMC} SSF and the resampled ones with
  \texttt{KDSource} code for different ranges of energy ($\Delta E$)
  and polar angle direction ($\Delta \mu$), indicated in the first two
  columns (``Total'' indicating intergation over the whole variable
  range). The first value for $J_{\hat{\mathbf{z}}}^{+}$ corresponds
  to the integral partial current, i.e. integrated over all the phase
  space variables domain, and it is equal for both cases due to the
  normalization chosen.}
\label{table:J1-integral}
\end{table}

\begin{figure}[H]
    \centering
    \subfigure
    {
        \label{subfig:J2-lethargy}
        \includegraphics[width=0.47\textwidth]{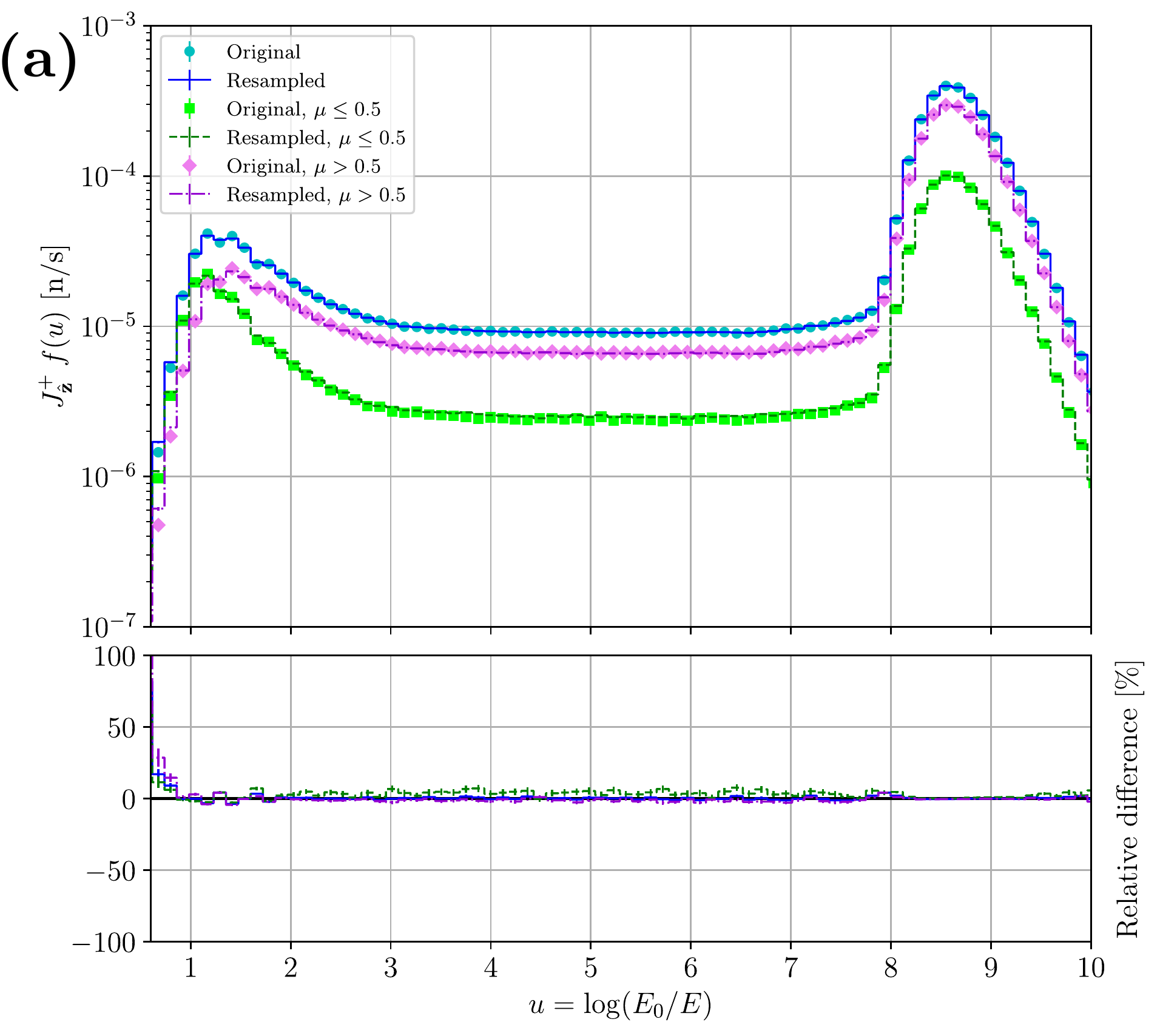}
    }
    \subfigure
    {
        \label{subfig:J2-position}
        \includegraphics[width=0.47\textwidth]{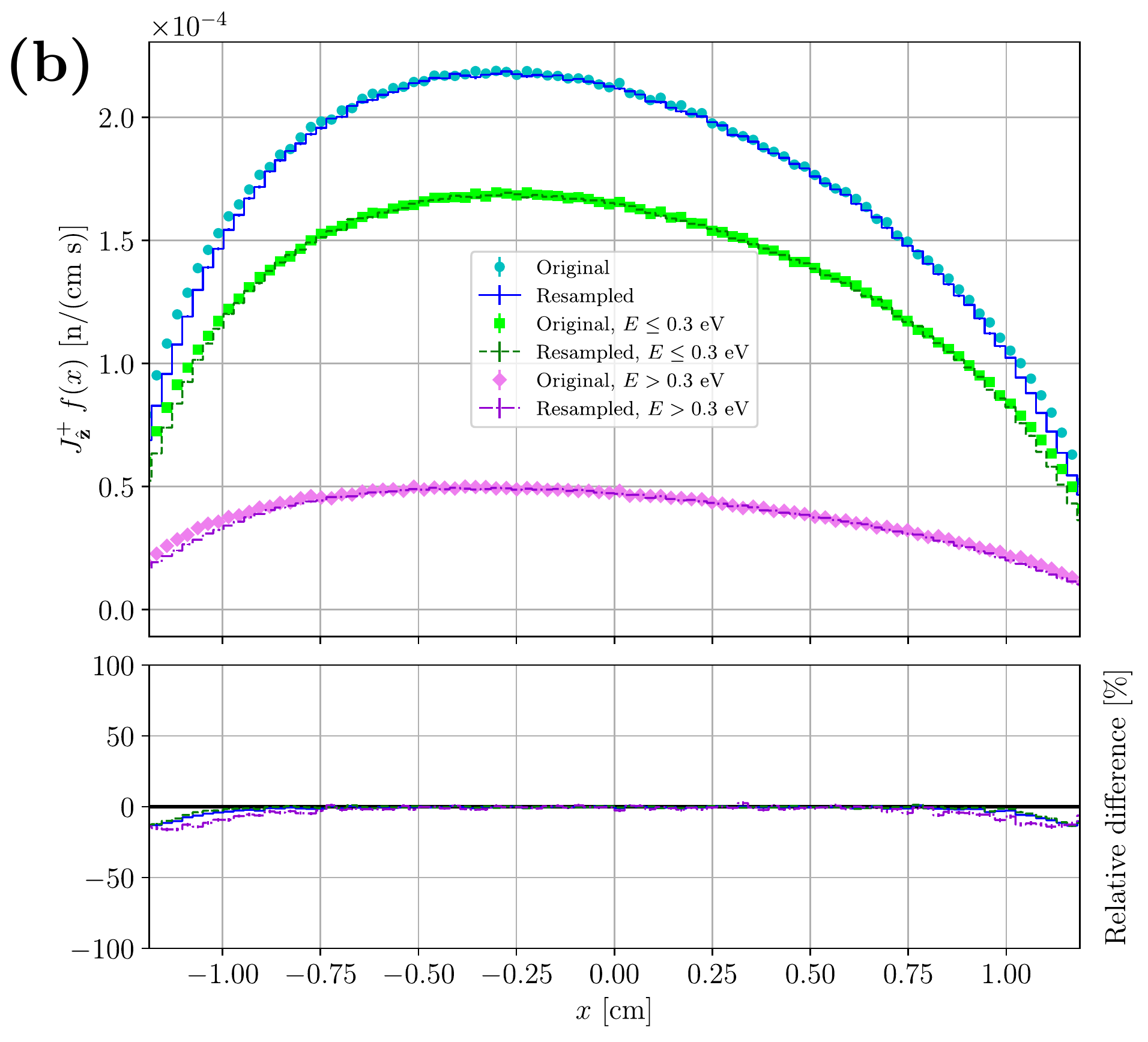}
    }
    
    \subfigure
    {
        \label{subfig:J2-polar}
        \includegraphics[width=0.47\textwidth]{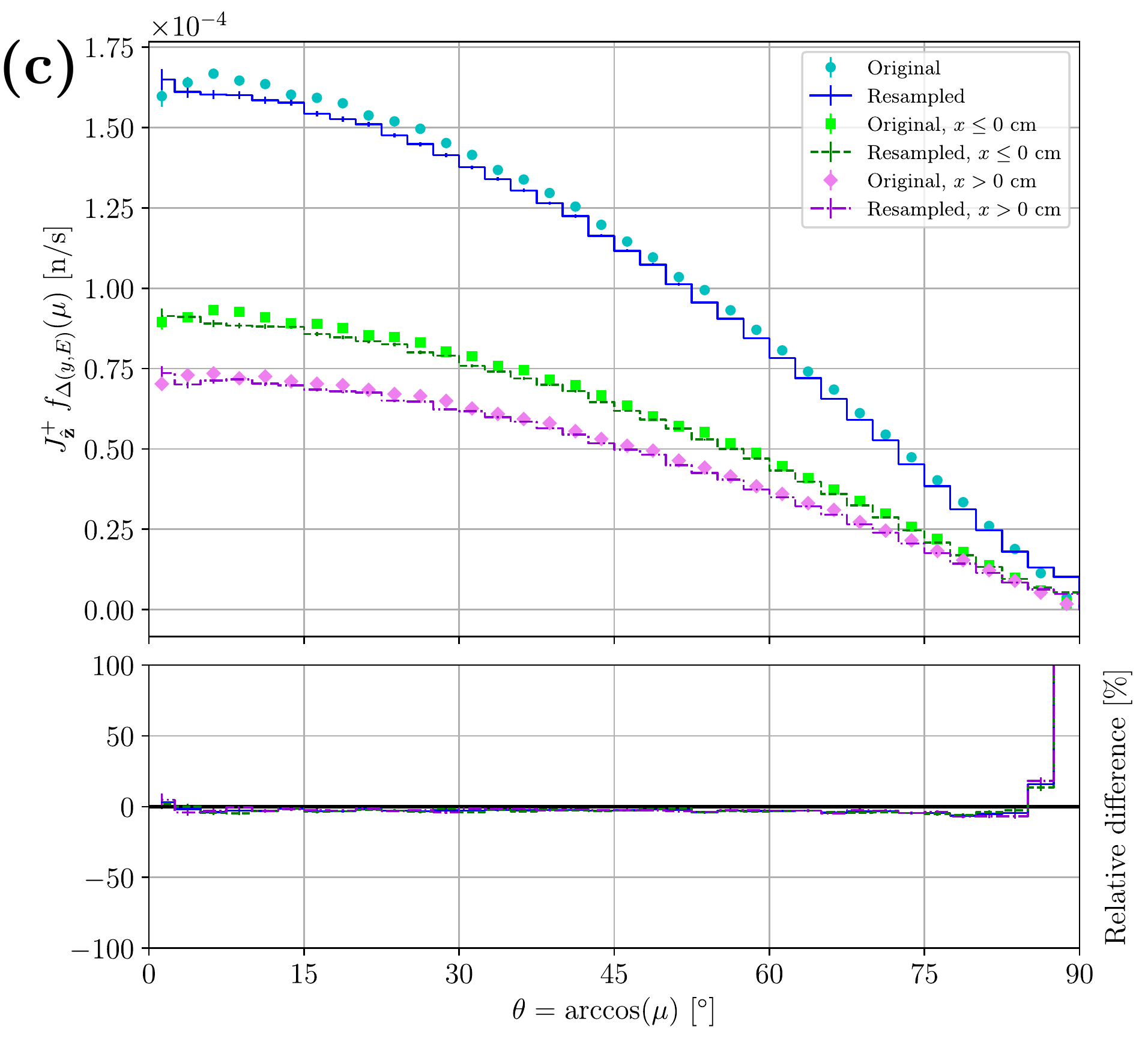}
    }
    \subfigure
    {
        \label{subfig:J2-azimuthal}
        \includegraphics[width=0.47\textwidth]{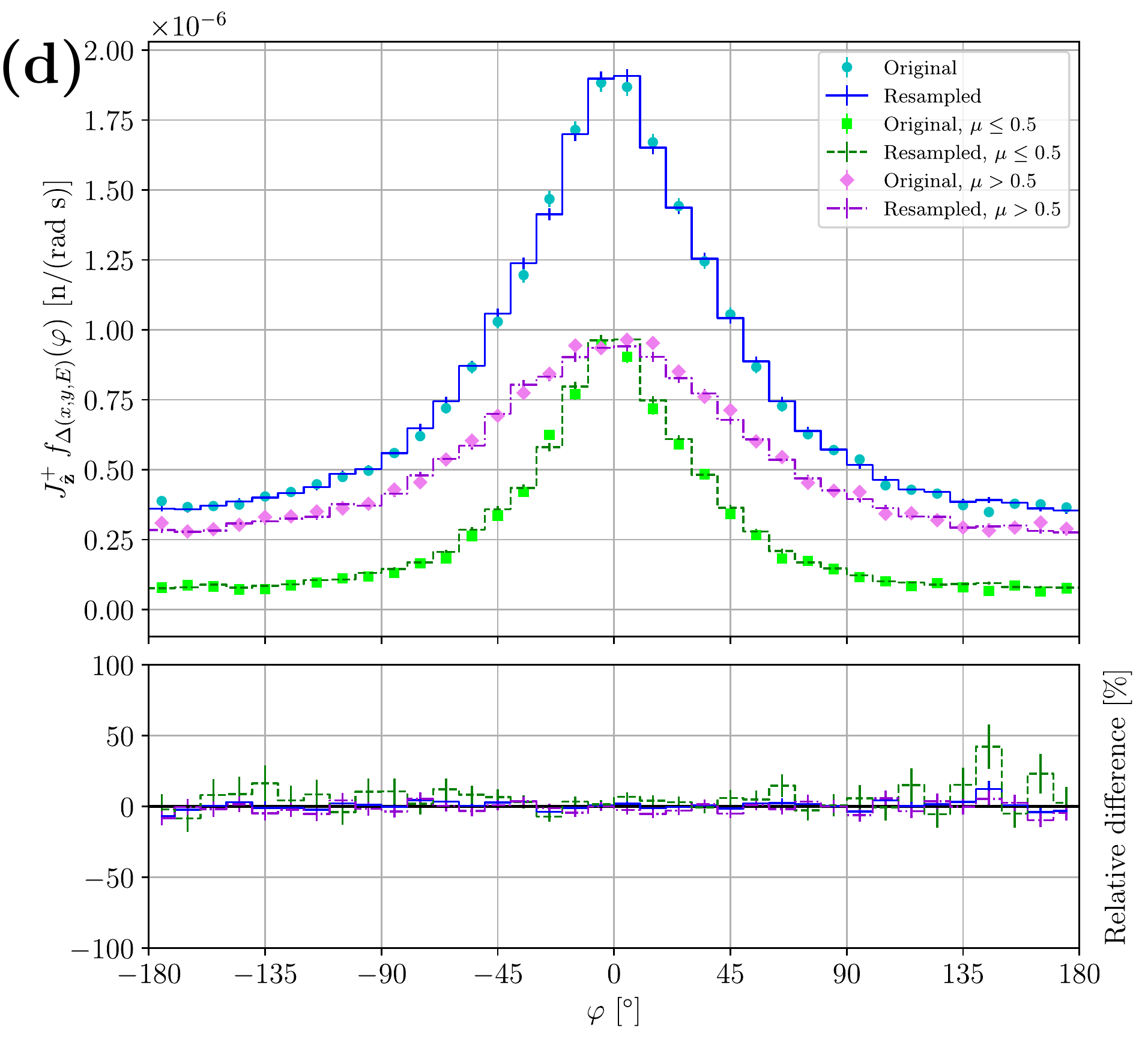}
    }
    \caption{Partial current distributions of the neutrons emitted by
      the source obtained from the \texttt{OpenMC} SSF and the
      resampled ones with \texttt{KDSource} code. In all the lower
      frames relative differences are shown for comparison. (a)
      Lethargy distribution, where $x$ and $y\in(-\infty, \infty)$ and
      $\varphi\in [0,2\pi$]. (b) Position distribution, where
      $y\in(-\infty, \infty)$, $\mu\in[0,1]$ and
      $\varphi\in[0,2\pi]$. (c) Polar angle direction distribution,
      where $y\in[-0.25, 0.25]$ cm, $E\in[0,0.3]$ eV and
      $\varphi\in[0,2\pi]$. (d) Azimuthal angle direction
      distribution, where $x$ and $y\in[-0.25, 0.25]$ cm and
      $E\in[0.3,\infty)$ eV.}
    \label{fig:J2-comparison}
\end{figure}

Fig. \ref{fig:J2-comparison} shows the comparison between the original
distributions of the SSF variables and those obtained from the
resampling of the estimated distribution (case 2). Also, for each
variable the relative difference between the resampled histogram and
the \texttt{OpenMC} particle lists histogram is shown. To verify the
correlation between the variables, different domains of interest were
used for each distribution.  The subscript $\Delta(...)$  of
the PDF function in ordinates indicates that a partial integration has
been performed in the variables in parentheses.  As this figure shows,
the largest differences between original SSF and resampled
distributions are generally less than 20\%.  In particular, the
relative difference increases when the distributions tends to zero at
the domain borders, attributable this also to a high second derivative
and to the lower statistics in those regions of the phase space. This
figure also shows that the correlations for the lethargy, position,
direction are respected.

\begin{figure}[H]
    \centering
    \includegraphics[width=0.8\textwidth]{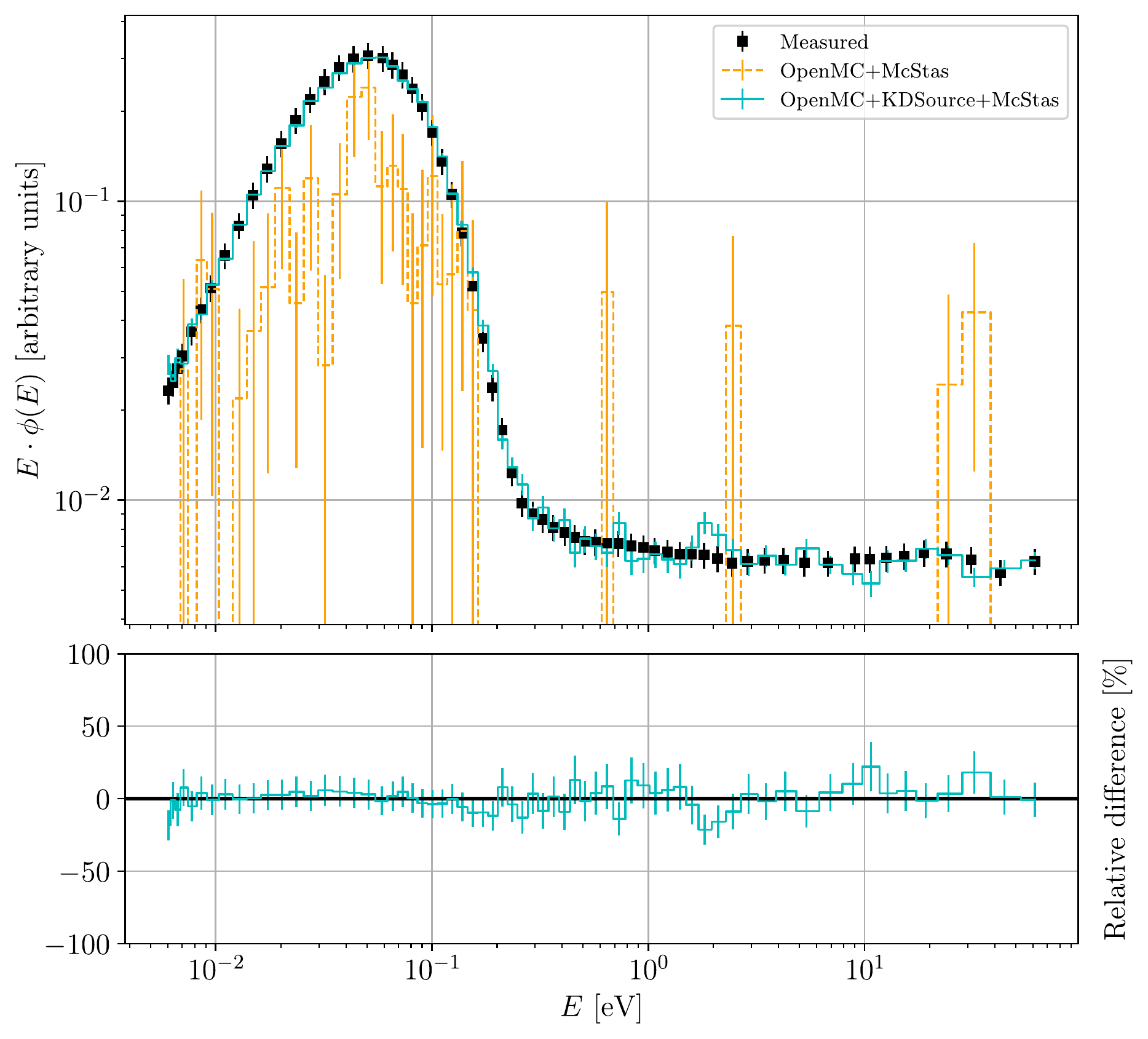}
    \caption{Comparison between the experimental spectrum (black
      squares) and those calculated by a standard \texttt{OpenMC}
      simulation (orange bars with errors) and by combining
      \texttt{OpenMC}, \texttt{McStas} and \texttt{KDSource}. Relative
      differences between experimental measurements and
      \texttt{KDSource} results are shown for comparison in the lower
      frame.}
    \label{fig:phiE-measured}
\end{figure}

Finally, in Fig. \ref{fig:phiE-measured}, the experimental spectra
\citep{abbate1976neutron} is compared with the ones obtained with
\texttt{McStas} using the original SSF and the resampled one with
\texttt{KDSource} (case 2). The number of particles sampled with the
distribution estimated by \texttt{KDSource} was 1000 times larger than
that of the original \texttt{OpenMC} SSF.  The agreement between the
experimental and the sampled spectra is remarkable, which is evidenced
by the fact that the relative bin-to-bin difference values are always
below 20\%, as shown in the lower frame. Figure
\ref{fig:phiE-measured} also shows that in a common Monte Carlo
simulation, there are energies that are never recorded due to the
finite number of particles, a situation that is overcome with the use
of \texttt{KDSource}.

\newpage

\subsection{Application}
\label{sec:application}

\begin{figure}[H]
    \centering
    \subfigure{\label{fig:dose_a}\includegraphics[height=300pt]{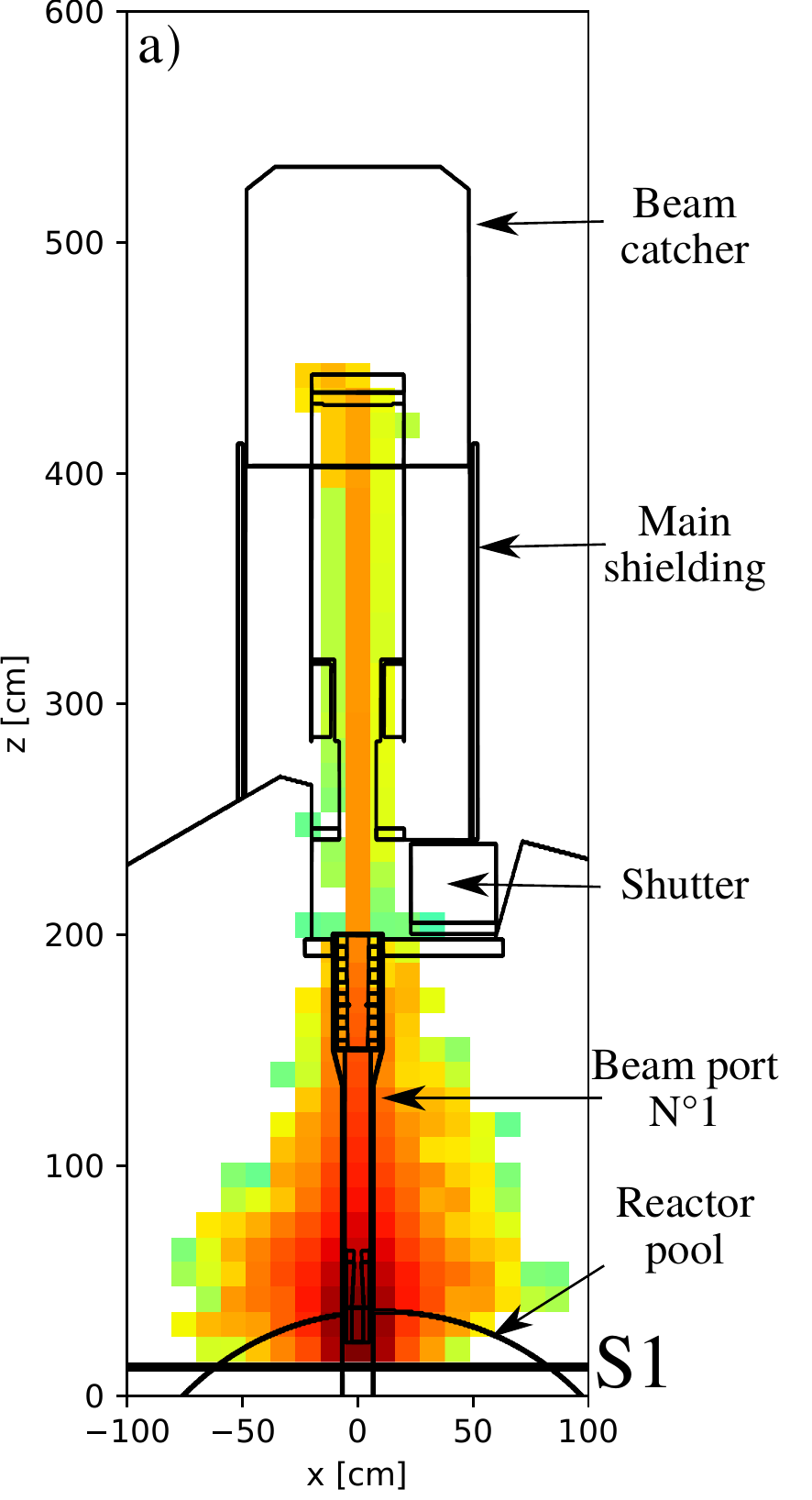}}
    \subfigure{\label{fig:dose_b}\includegraphics[height=300pt]{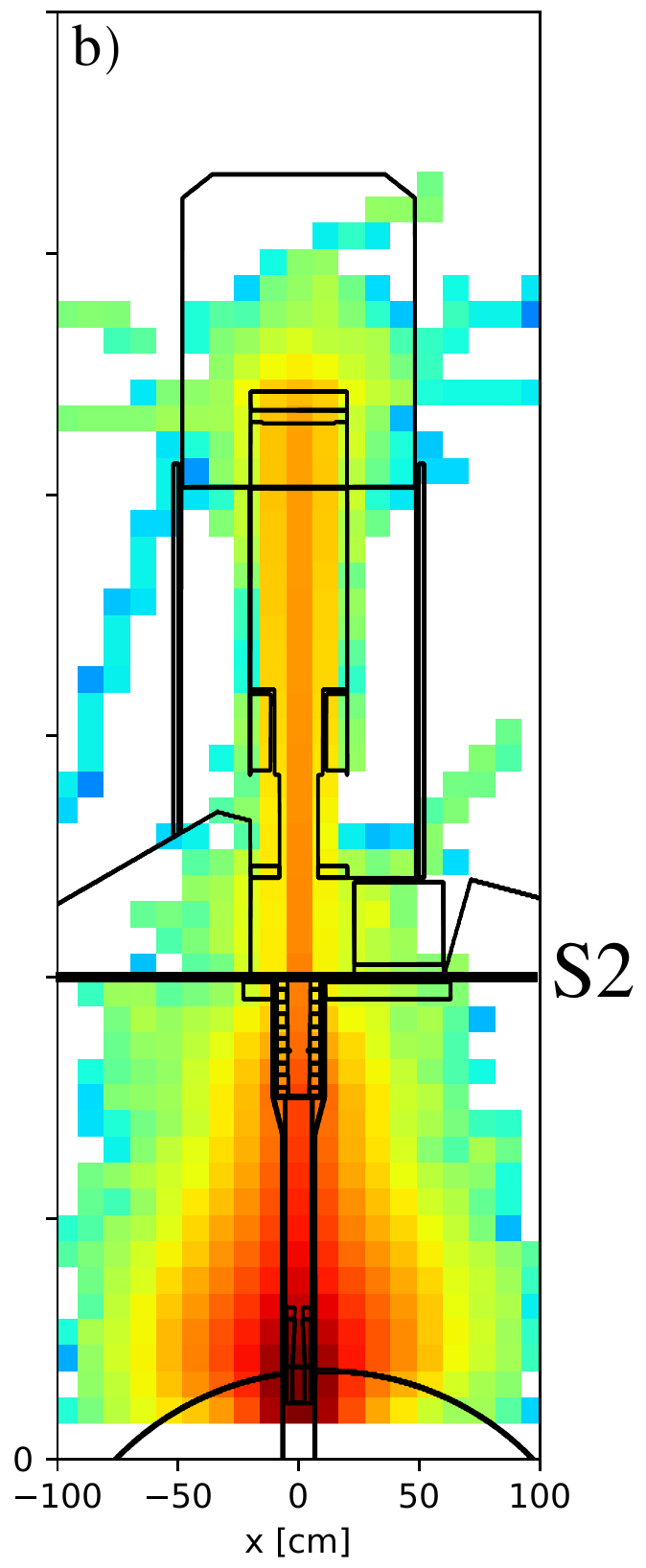}}
    \subfigure{\label{fig:dose_c}\includegraphics[height=300pt]{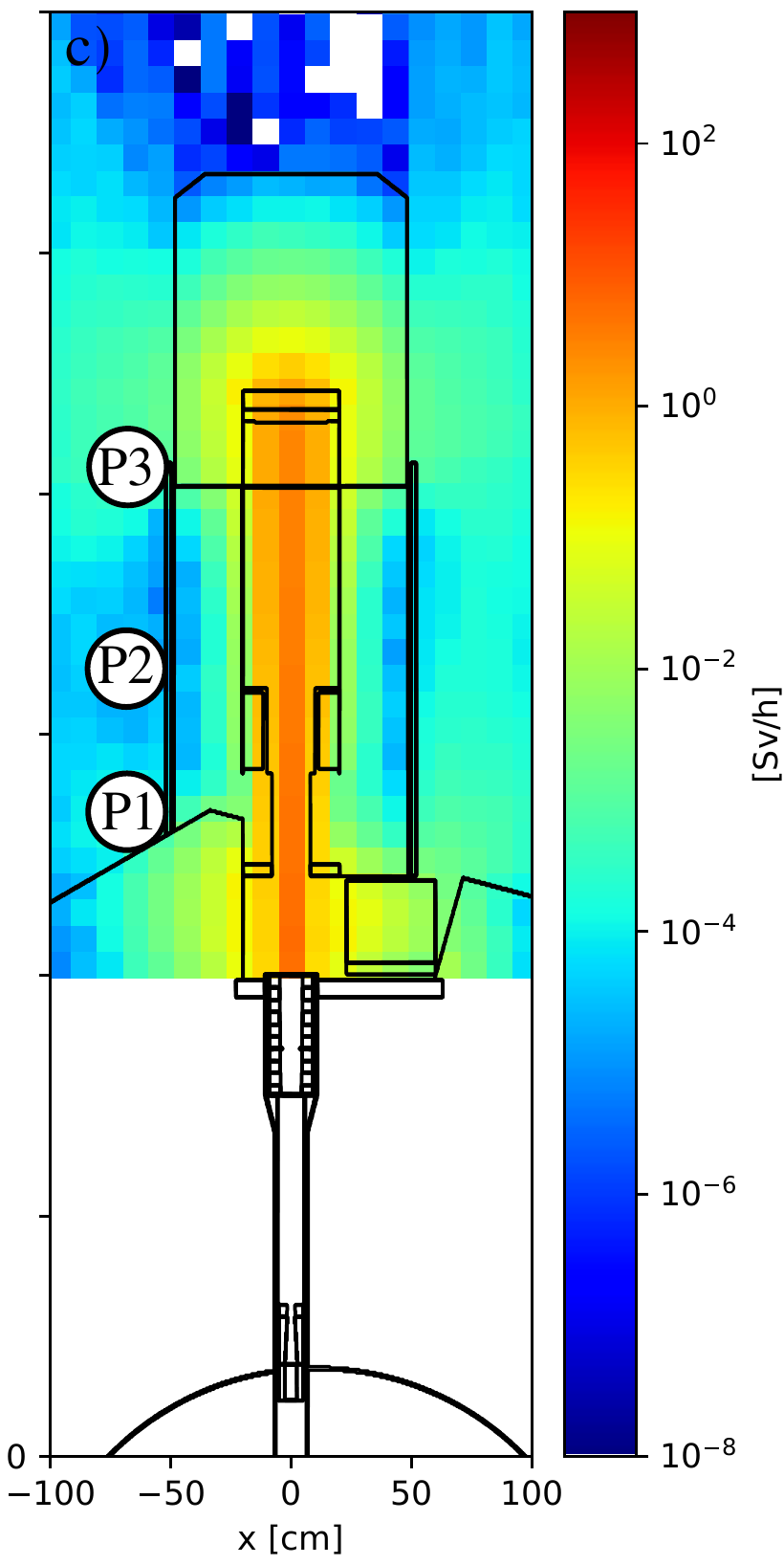}}
    \caption{Neutron dose rate map of RA6 reactor's conduit that leads
      to the neutrography facility. Samples from Kernel density
      estimations of the dose rate map at different distances (S1 and
      S2) from the source are depicted. (a) Core simulation.
      source. (b) First KDE source. (c) Second KDE source.}
    \label{fig:neutron_dose}
  \end{figure}

As an application case of the method presented here, we show its use
in transport calculations using the OpenMC code, employed in the
design of an instrument using a neutron beam of an experimental
reactor.  The case studied is the neutrography facility of the RA-6
reactor of the Bariloche Atomic Center (Argentina).  The goal is to
map the neutron flux inside the instrument housing to perform the
dosimetry evaluation. A first simulation using \texttt{OpenMC}
starting from the reactor core reveals the difficulty in obtaining an
adequate dose rate statistic (see Fig. \ref{fig:dose_a}). To improve
the calculations, \texttt{KDSource} was used to evaluate the source
defined at surface S1 of the reactor neutron beam port entrance.
For this, the simulation was employed to generate a track list of
$N=5 \times 10 ^ 4$ neutrons at surface S1.

The application of \texttt{KDSource} served to estimate the joint
PDF of the neutron phase vector components.  This allowed to perform
a second simulation with $N= 5 \times 10 ^ 6$ neutrons generated
from the estimated distribution at S1. The dose rate map thus
obtained can be seen in Fig. \ref{fig:dose_b}. Although the map has
improved in the vicinity of the beam port with respect to that
of Fig.  \ref{fig:dose_a}, it has not yet converged in the
region downstream the tube exit (referred as surface S2 in
fig. \ref{fig:dose_b}).
Since our goal is to determine a dose rate map in the surroundings
of the instrument, reliable enough to be compared with measured dose
rates, we generated a new particle list at surface S2 consisting in
$N=1 \times 10 ^ 4$ neutrons registered from the previous simulation
of the KDE source at S1, and performed a third simulation.  The new
source distribution at surface S2 thus generated keeps the
correlations between variables of the phase space vector. The dose
rate map was obtained by resampling $1 \times 10 ^ 7$ neutrons from
S2.  The result is shown in fig. \ref{fig:dose_c}. Thus, an
efficient propagation of the source distribution (with the
corresponding correlations between variables) at a greater distance
along the beamline can be achieved, allowing a better estimation of
the dose rate map far away of the reactor core.

Finally, the simulation’s results of the dose rate map shown in fig.
\ref{fig:dose_c} were contrasted with experimental measurements taken
at the points P1, P2, and P3 near the main shielding structure (see
fig. \ref{fig:dose_c}). The measurements were carried out for a
radiological survey of the facility, in order to detect surrounding
locations that require protected actions such as improved shieldings,
barrier placement and standards of use. Measured values of the
neutron dose rate and values obtained by simulation are shown in
table \ref{table_dosis} for three points.

\begin{table}[!htpb]
\centering
\begin{tabular}{|c|c|c|c|}
\hline
\multirow{2}{*}{Point} & \multicolumn{2}{c|}{Neutron dose rate $[\mu Sv/h]$} & \multicolumn{1}{l|}{\multirow{2}{*}{Relative Error {[}\%{]}}} \\ \cline{2-3}
                       & Experimental          & Calculation         & \multicolumn{1}{l|}{}        \\ \hline
P1                     & 30                    & 46                  & 53                           \\ \hline
P2                     & 25                    & 29                  & 16                           \\ \hline
P3                     & 120                   & 125                 & 4                            \\ \hline
\end{tabular}
\caption{Neutron dose rate comparison of RA6 reactor's beamport.}
\label{table_dosis}
\end{table}

\noindent As can be seen, the values obtained from the simulation are
consistent with the experimental values, and the consistency increases
with the distance to the surface S2. It is also interesting to note
that the dose rate at point P3, which is the farthest from the source
is higher than at points P1 and P2 due to the backscattering in the
beam catcher. This feature is being correctly represented in the
simulation because KDE is allowing to sample a large amount of
neutrons from the surface S2, allowing the map to converge adequately.

\section{Discussion and conclusions}
\label{sec:conclusions}

In this work we presented \texttt{KDSource}, a new tool for generating
source distributions based on the kernel density estimation (KDE)
method.  The implemented algorithm automatically optimizes a KDE model
based on a list of particles recorded at an intermediate position of a
Monte Carlo simulation, normally the input of a radiation beam, thus
estimating the current density distribution there.  As already shown
in previous similar implementations
\cite{tyagi2006proposed,banerjee2010kernel,burke2016kernel}, this
technique clearly benefits the tally computation in beam simulations
compared to direct use of the particle list.  The tool aims to
simplify the application of the KDE method with respect to the
mentioned implementations, reducing the required user expertise and the
engineering hours, by providing an abstract user interface.

The properties and usefulness of the \texttt{KDSource} tool were
studied in three different problems. The first one, referred as
verification, allowed to check some basic properties of the
\texttt{KDSource} algorithm, such as the convergence of the estimated
distribution to the real density, and the adequate correlation
modeling. This was possible because, in this analytical problem, the
real density was known. The second problem provided a demonstration of
\texttt{KDSource}'s benefits in a real Monte Carlo simulation,
significantly increasing the statistics at the end of a neutron
beam. In this case validity of the KDE source was verified by
comparing it with the track source at the points where both had
sufficient statistics, and with the experimental results. Finally, in
the application problem the \texttt{KDSource} algorithm was used to
solve an actual problem of interest in a research reactor neutron
beam, allowing shielding calculations far away from the core, with
reasonable agreement with experimental results.

From a more theoretical point of view, the advantages of the
\texttt{KDSource} algorithm, observed in the last two problems, can be
understood as follows. In a Monte Carlo simulation, uncertainties in
the sources distribution result in systematic errors in the computed
tallies. Sensitivity is maximal in vacuum propagation problems, where
flux maps depend strongly on the source, and not in the properties of
the medium (as in moderation problems).  In such cases,
\texttt{KDSource} improves the simulation by reducing the variances of
the tallies caused by the statistical fluctuations of the particle
list, which could be noticeable with a tracks source.
Therefore, the \texttt{KDSource} tool is expected to be useful on
applications with strong vacuum propagation, specially beam
simulations and associated shielding calculations. Furthermore, if the
tracks list on which the KDE model is fitted is independent of the
downstream problem, which is common in beams, the KDE source can be
reused with different downstream configurations, increasing the
benefits of the methodology.

In future works the capabilities of \texttt{KDSource} are expected to
be extended in several ways. This includes the possibility of choosing
between different kernel functions (Epanechnikov, Gaussian,
triangular, Tophat, Exponential, etc.), the inclusion of other
possible bandwidth optimization methods, such as Least-Squares
Cross-Validation, and further source optimization techniques like
source-biasing. It is also important to complete the documentation of
all the functionalities of \texttt{KDSource}, following the usual
software standards. In addition, there are other application problems
that can be explored, such as the calculation of coupled optics
shielding around neutron guides, and the modeling of neutron
activation sources. Finally, regarding the statistical properties of
the implemented algorithm, a practical procedure to determine the
minimum number of training particles required to properly fit a
KDE model is left as a possibility for future improvements.

\section{Acknowledgements}
\label{sec:Ack}

Funding from Universidad Nacional de Cuyo (Project 06-C563), ANPCYT
(Agencia Nacional de Promoción Científica y Tecnológica - Argentina)
(PICT 2019 - 02665) and CNEA (Argentine National Commission of Atomic
Energy) are gratefully acknowledged.

\end{document}